# Knowledge Manifold Optimization for Scientific Discovery

*Multi-Objective Pareto Search for Supported Knowledge Gaps*

**Shigeru Obayashi*[1], Yoshiaki Kawagoe[2,3], Kazuhiko Komatsu[3], Tomonaga Okabe[2,4]**

[1]Institute of Fluid Science, Tohoku University, 2-1-1 Katahira, Aoba-ku, Sendai, Miyagi 980-8577, Japan

[2]Department of Aerospace Engineering, Tohoku University, 6-6-01, Aoba, Aramaki, Aoba-ku, Sendai, Miyagi 980-8579, Japan

[3]Research Center for Green X-Tech, Tohoku University, 6-6-11, Aoba, Aramaki, Aoba-ku, Sendai, Miyagi 980-8579, Japan

[4]Department of Materials Science and Engineering, University of Washington, BOX 352120, Seattle, WA 98195-1750, USA

*Email: s.obayashi@tohoku.ac.jp

**Abstract**

The rapid expansion and specialization of scientific literature make it increasingly difficult to identify not only what is known, but where new and still supportable knowledge may emerge. We introduce Knowledge Manifold Optimization (KMO), a discovery-oriented extension of the Knowledge Manifold framework that treats scientific discovery as a multi-objective search over a continuous, literature-derived knowledge space. Candidate locations are evaluated using four complementary objectives: persistent knowledge-regime boundary strength, band-pass white-space novelty, semantic-gradient magnitude, and evidential support. Rather than collapsing these criteria into a single weighted score, KMO identifies Pareto-optimal Supported Knowledge Gaps, defined as corpus-relative unoccupied regions that remain sufficiently supported by surrounding literature. The method is demonstrated on a 99-paper corpus in aerodynamics and aerodynamic design. On a 21 × 21 grid, exact Pareto analysis produced 19 non-dominated points, of which 16 formed the practical front after excluding points with zero white-space score. These solutions exhibited several distinct discovery modes, including methodology bridges, aero-structural topology/FSI integration, functional aerodynamic-device integration, and rear-wake/vehicle flow-control integration. A distance-maximization baseline preferentially selected extrapolative regions with substantially lower support and reliability, demonstrating why emptiness alone is insufficient. Representative Pareto locations were converted into auditable evidence packages and then verbalized by a constrained large language model into candidate research hypotheses. The results show that scientific discovery can be formulated as structured multi-objective search in an explicit knowledge space, while keeping novelty corpus-relative and separating numerical discovery from downstream language generation.

## 1. Introduction

The rapid growth and increasing specialization of scientific literature have made it progressively more difficult for researchers not only to retrieve relevant information but also to determine where promising opportunities for new knowledge may lie. Search engines, bibliographic databases, citation networks, and semantic retrieval systems are highly effective at locating existing publications. Scientific discovery, however, requires a different capability: identifying relationships, transitions, and gaps among existing bodies of knowledge that may suggest research directions not yet explicitly represented in the literature.

The idea that scientific knowledge can be represented spatially has a long history. Similarity relationships can be represented as distances in low-dimensional spaces [1], and self-organizing maps (SOMs) provide topology-preserving representations of high-dimensional data [2]. Conceptual-space theory further interprets concepts themselves as geometrical regions [3,4]. In science mapping, bibliometric and text-based methods have subsequently been developed to visualize the structure of research fields, including VOSviewer, CiteSpace, bibliometrix, and document-level self-organizing maps [10–17]. Modern semantic representations and manifold-learning methods, including Sentence-BERT, Isomap, t-SNE, and UMAP, have further increased the ability to organize high-dimensional document spaces into interpretable low-dimensional representations [5–8].

These approaches have greatly advanced the visualization and structural analysis of scientific literature, but a conventional science map is fundamentally composed of observed documents or other discrete entities. The space between those observations generally has no explicit semantic representation, uncertainty, gradient, or evidential support. Consequently, an apparently empty region of a literature map cannot, by geometry alone, be distinguished from a meaningful research opportunity or an unsupported extrapolation. This distinction becomes essential when the purpose of the map shifts from retrospective visualization to prospective scientific discovery.

A complementary intellectual lineage is literature-based discovery. Swanson demonstrated that meaningful scientific relationships can remain undiscovered because relevant pieces of knowledge are distributed across disconnected literatures [24]. Chen et al. subsequently developed a structural view of scientific discovery in which potentially transformative findings can emerge by bridging disparate patches of knowledge [25]. More recent studies have quantified novelty through unusual combinations of scientific topics [26] and shown that surprising combinations of research contents and contexts can be associated with scientific impact [27]. These studies suggest that scientific opportunity may be concentrated neither in densely populated regions of established knowledge nor in arbitrarily remote regions, but at locations where previously separated knowledge can be combined in a meaningful and sufficiently supported manner.

To make such locations computationally accessible, we previously introduced the Knowledge Manifold, in which a corpus of scientific documents is treated not merely as a set of points but as a continuous knowledge space. The preliminary formulation represented documents using character n-gram TF–IDF vectors, embedded them on a two-dimensional map, reconstructed semantic content at arbitrary unoccupied locations through Smoothed Particle Hydrodynamics (SPH), quantified local uncertainty using Gaussian Process Regression (GPR), and defined knowledge gradients and geodesic paths on an induced Riemannian geometry [28]. That initial study demonstrated the concept using a 20-paper corpus in composite materials and aerospace structural mechanics and showed that unoccupied locations could be associated with interpolated semantic content and virtual research descriptions.

The framework was subsequently extended into a continuous, differentiable, uncertainty-aware, and auditable representation of scientific literature [29]. The extended formulation uses a fixed bounded chart, dual $\ell_2/\ell_1$ representations, SPH interpolation, GPR uncertainty, SPH-induced Riemannian and Fisher–Rao geometries, geodesic navigation, and evidence-constrained language-model verbalization. Its central distinction from conventional science mapping is that every location on the map—not only the locations occupied by documents—can carry semantic content, uncertainty, and supporting evidence. The framework was evaluated on three 100-paper corpora with machine-checked validation procedures, establishing the Knowledge Manifold primarily as a framework for representation, analysis, and navigation of the structured space between scientific documents.

The present study asks the next question: given a Knowledge Manifold, where should one search for new scientific knowledge? This question cannot be answered simply by locating the largest empty region. Maximizing the distance from existing documents would eventually favor locations that are poorly supported by the corpus and dominated by extrapolation. Conversely, searching only densely populated regions would favor incremental extensions of established work. A scientifically interesting location may instead lie near the interface of distinct knowledge regimes, exhibit strong semantic variation, remain sufficiently separated from existing publications to permit novelty, and at the same time retain enough surrounding evidence to support a plausible research hypothesis. Scientific discovery on a Knowledge Manifold is therefore naturally a multi-objective problem.

This view has a direct precedent in our earlier work on multi-objective engineering design. In aerodynamic and multidisciplinary design optimization, we have treated the Pareto-optimal solution set not merely as a collection of candidate optimum designs but as a source of design knowledge. Pareto solutions were projected and analyzed using SOMs to reveal trade-offs, relationships among design variables and objectives, and characteristic regions of the design space [30–32]. This line of research led to the concept of Multi-Objective Design Exploration (MODE), in which optimization, visualization, and data mining are combined so that the structure of a design space itself becomes an object of scientific and engineering inquiry [33]. In this perspective, the purpose of multi-objective optimization is not necessarily to identify one preferred solution; the distribution and organization of the Pareto set can themselves reveal knowledge that is not apparent from individual solutions.

The present work extends this principle from design-space knowledge discovery to scientific knowledge-space discovery. In conventional multi-objective design exploration, a design vector is evaluated through multiple competing engineering objectives, and the Pareto set reveals distinct compromise designs. Here, a location on the Knowledge Manifold is instead evaluated through multiple measures of its potential for scientific discovery. Thus, design variables → engineering objectives → Pareto designs → design knowledge is generalized to knowledge coordinates → discovery objectives → Pareto knowledge gaps → candidate scientific knowledge. This connection provides the conceptual basis for what we call Knowledge Manifold Optimization.

Recent AI-based systems have approached open-ended scientific discovery from another direction. SciMON searches for literature-grounded research ideas while explicitly encouraging novelty [35], ResearchAgent iteratively generates research ideas using scientific literature and LLM agents [36], and The AI Scientist extends automation to broader parts of the research cycle [37]. AUTODISCOVERY formulates open-ended discovery as a search over hypotheses, using Bayesian surprise as a reward and Monte Carlo Tree Search to balance exploration and exploitation [38]. Mechanism-centric retrieval has further shown that scientific opportunities may arise from structurally analogous mechanisms even when the corresponding domains are semantically distant [39]. These studies establish that scientific discovery can be treated computationally as a search problem rather than solely as unconstrained language generation.

Knowledge Manifold Optimization differs in the object being searched. Rather than allowing a language model to generate and rank hypotheses directly, we first construct an explicit, continuous knowledge space from the scientific corpus and then optimize locations within that space. The search is therefore constrained by the geometry and evidence structure of the literature before language generation takes place. This provides a complementary route to scientific discovery: the LLM is not asked to decide where a discovery should occur, but only to interpret and verbalize evidence associated with locations identified by the numerical search. This division of labor follows the Knowledge Manifold principle that deterministic geometric analysis should precede evidence-constrained verbalization.

We formulate the discovery problem using four complementary objectives, representing persistent knowledge-regime boundaries, band-pass white-space novelty, semantic-gradient magnitude, and evidential support. Because there is no unique scientific basis for collapsing these competing criteria into a single scalar score, we use Pareto dominance to preserve distinct forms of discovery opportunity. The resulting non-dominated points represent qualitatively different discovery regimes rather than a single best location. We refer to promising unoccupied locations that are novel relative to the corpus while remaining supported by neighboring evidence as Supported Knowledge Gaps.

The proposed approach is demonstrated using a corpus of 99 papers in aerodynamics and aerodynamic design. Multiple resolutions of the same fixed Knowledge Manifold are used to distinguish broad research regimes from finer subfields, while global and locally adaptive interpolation are compared to assess whether apparent boundaries persist under changes in resolution. Discovery objectives are evaluated throughout the bounded manifold, and the discrete non-dominated set is extracted directly rather than inferred from a single weighted score. The Pareto solutions are subsequently interpreted as distinct discovery archetypes, and representative locations are converted into evidence packages from which candidate research directions are verbalized.

The contribution of this study is therefore not a new method for constructing a science map; the construction and geometric interpretation of the Knowledge Manifold have been addressed in our previous work [28,29]. Instead, the present study introduces an optimization layer for scientific discovery on the Knowledge Manifold and connects two lines of our previous research: Pareto/SOM-based extraction of knowledge from engineering design spaces and geometric representation of scientific literature. Knowledge Manifold Optimization treats scientific discovery not only as the generation of hypotheses, but as a structured search for where new, yet evidentially supported, knowledge may exist. The conceptual relation between design-space knowledge discovery and the present framework is summarized in Fig. 1.

$$\text{maximize over } x \in \Omega: \ [\, P_{\text{boundary}}(x),\ W(x),\ G(x),\ S(x) \,]$$

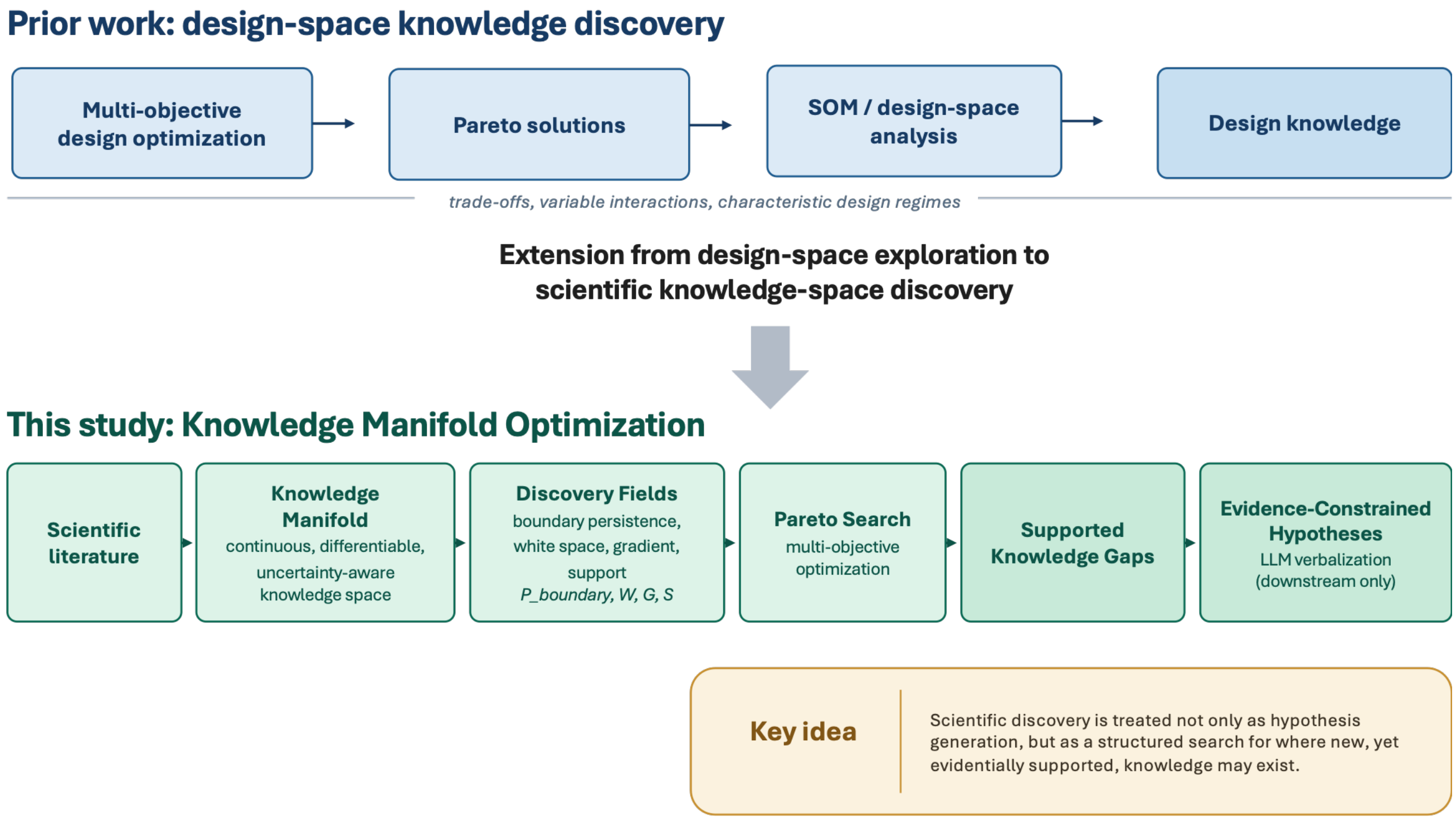

Figure 1. Conceptual transition from multi-objective design exploration to Knowledge Manifold Optimization. Prior work extracts design knowledge from Pareto solution sets through SOM-based design-space analysis; the present framework transfers this principle to a literature-derived knowledge space and searches for Pareto-optimal Supported Knowledge Gaps before evidence-constrained hypothesis verbalization.

## 2. Problem Formulation and Knowledge Manifold Optimization Framework

### 2.1 From knowledge representation to knowledge discovery

The construction of the Knowledge Manifold itself follows our previous work [28,29]. The present study does not modify the document representation or optimize the manifold coordinates. Instead, it takes a fixed Knowledge Manifold as the computational domain and introduces a discovery-oriented optimization layer for identifying scientifically promising locations within that space.

Let $\Omega = [-1,1]^2$ denote the bounded two-dimensional chart of the Knowledge Manifold, and let $D = \{(x_i,z_i)\}_{i=1}^N$ denote the corpus, where $x_i \in \Omega$ is the fixed coordinate of document i and $z_i$ is its high-dimensional semantic representation. As described in [28,29], interpolation over the fixed document coordinates defines a continuous semantic field v(x), together with local gradients, document contributions, and model-based uncertainty.

This construction changes the scientific-discovery problem fundamentally. In a conventional literature map, the principal objects are the observed documents. In a Knowledge Manifold, any location x—including a location not occupied by an existing document—can be queried. The question is therefore no longer only what knowledge exists at this location, but also which locations should be examined preferentially for possible new knowledge. We formulate the latter as an optimization problem over the continuous knowledge space.

$$v(x), x \in \Omega \quad (1)$$

### 2.2 Why an empty region is not necessarily a discovery opportunity

A simple approach to knowledge-gap detection would be to maximize the distance from existing documents. Let $d_{nn}(x)$ be the distance from a query location to the nearest observed document. Maximizing $d_{nn}$ identifies the emptiest portion of the map, but it does not distinguish between a meaningful knowledge gap and unsupported extrapolation.

Two limiting cases illustrate the problem. If a candidate is extremely close to an existing document, the candidate is well supported but likely to represent only an incremental extension of established work. Conversely, if the candidate is extremely remote from all observed documents, its apparent novelty may simply reflect insufficient evidence. The discovery-relevant region is therefore expected to occupy an intermediate regime between redundancy and extrapolation.

Distance alone is also insufficient because scientific opportunity can arise from the structure of the neighboring knowledge. In particular, a location at which several knowledge regimes coexist may represent an opportunity to combine concepts that have previously developed separately. Similarly, a location where semantic content changes rapidly may indicate an active transition between research themes. Finally, such a candidate should be supported by a sufficiently broad neighborhood rather than by a small number of isolated papers. We therefore treat discovery potential as inherently multi-dimensional.

$$d_{\mathrm{nn}}(x) = \min_i \|x - x_i\| \quad (2)$$

## 2.3 Supported Knowledge Gaps

We introduce the term Supported Knowledge Gap to distinguish a discovery-oriented knowledge gap from an arbitrary empty region. A Supported Knowledge Gap is an unoccupied or weakly occupied location of the Knowledge Manifold that simultaneously exhibits four desirable characteristics: boundary proximity, appropriate novelty, semantic change, and evidential support.

These properties are deliberately not collapsed into a single definition of novelty. A candidate may be scientifically interesting because it joins two established regimes, because it lies within a sparsely occupied but still supported region, because the local semantic field changes rapidly, or because several of these mechanisms act simultaneously. Accordingly, a Supported Knowledge Gap is operationally identified not by exceeding one threshold, but by its position relative to other candidate locations in a multi-objective discovery space.

The distinction is also important epistemically. The term gap in this paper is corpus-relative. A location identified as a Supported Knowledge Gap is absent or weakly represented in the analyzed corpus; it is not thereby proven to be globally novel in the complete scientific literature. External literature search and expert evaluation remain required before a candidate can be regarded as a genuinely novel scientific contribution.

## 2.4 Overall framework

Knowledge Manifold Optimization consists of five stages. The first stage is inherited from the Knowledge Manifold framework [28,29]. The second stage constructs continuous scalar fields describing different aspects of discovery potential. The third treats these fields as competing objectives and extracts non-dominated locations. The fourth interprets the resulting Pareto set as a collection of different discovery opportunities rather than as a ranking with a unique optimum. Finally, selected locations are converted into evidence packages for linguistic interpretation.

Scientific corpus → fixed Knowledge Manifold → discovery fields → multi-objective Pareto search → Supported Knowledge Gaps → evidence-constrained hypotheses (3)

The framework therefore separates discovery location selection from hypothesis verbalization. All candidate locations are determined numerically before a language model is used.

## 2.5 Role of the language model

The language model does not evaluate the discovery objectives and does not select Pareto-optimal locations. Its role begins only after optimization.

For each selected location, the deterministic analysis exports an evidence package containing, among other quantities, the dominant semantic terms, neighboring-document contributions, document frequencies of relevant terms, local knowledge-regime mixture, effective support, and uncertainty. A language model is then instructed to produce a research description or virtual abstract while remaining constrained by this evidence.

For reproducibility, the complete researcher-supplied verbalization prompt and the post-generation audit protocol are reproduced in Appendix A. Two full generated examples, including the 200–280-word virtual abstracts, the exactly 45-word summaries, and the evidence-accounting lists, are provided in Appendix B.

This division of labor follows the auditable architecture introduced in the preceding Knowledge Manifold study [29], in which language generation is downstream of deterministic numerical analysis rather than the source of the underlying knowledge geometry.

$$\text{geometry and optimization} \rightarrow \text{candidate selection} \rightarrow \text{LLM verbalization} \quad (4)$$

## 3. Knowledge Manifold and Discovery Fields

### 3.1 Fixed Knowledge Manifold

Only those components of the Knowledge Manifold required for optimization are summarized here; the full construction and validation are given in [28,29].

Each scientific document is represented by a character n-gram TF–IDF vector. The $\ell_2$-normalized vectors define semantic-direction similarity, while an $\ell_1$-normalized representation provides a complementary probability-simplex interpretation [29]. Documents are assigned fixed coordinates $x_i = (x_i, y_i) \in \Omega$ on a bounded and gauge-fixed two-dimensional chart.

A kernel interpolation over these coordinates produces a continuous semantic vector field, where $w_i(x)$ are normalized SPH kernel weights. The previous Knowledge Manifold formulation shows that this lift from discrete documents to a continuous field makes gradients and routes between documents computable and supplements the field with GPR-based uncertainty.

Importantly, the two-dimensional representation is used operationally as a fixed exploration chart. We do not claim that the intrinsic dimension of the underlying scientific literature is exactly two.

$$v(x) = \text{normalize}(\textstyle\sum_i w_i(x) z_i) \quad (5)$$

### 3.2 Multi-resolution knowledge regimes

Scientific fields contain structure at more than one level of resolution. A broad partition may distinguish major methodological or application regimes, whereas a finer partition may distinguish specialized subfields.

We therefore partition the same fixed set of document coordinates at multiple clustering resolutions. Let $c_i^{(K)} \in \{1,\ldots,K\}$ denote the cluster assignment of document i at resolution K. For each cluster c, its continuous local probability field is obtained by interpolating the corresponding document indicator.

Two complementary smoothing strategies are used. A global field emphasizes broad, corpus-scale regime structure, whereas an adaptive field adjusts the local smoothing scale using a fixed nearest-neighbor rule and therefore gives greater weight to local structure.

In the present implementation, the broad discovery scale and the detailed reference scale are subsequently selected from the multi-resolution analysis described in the case study. Cluster labels themselves are not assumed to have invariant meanings across different values of K. Cross-resolution comparison is therefore performed through scalar boundary measures rather than by attempting to match cluster numbers directly.

$$p_c^{(K)}(x) = \textstyle\sum_i w_i(x) \mathbf{1}\left[c_i^{(K)} = c\right], \sum_c p_c^{(K)}(x) = 1 \quad (6)$$

### 3.3 Knowledge-regime boundary field

Locations where multiple knowledge regimes coexist are potential sites for cross-regime scientific integration. For a given clustering resolution, we first compute normalized cluster entropy. High entropy indicates that several regimes contribute simultaneously.

Entropy alone, however, does not distinguish a genuine competition between leading regimes from a diffuse mixture involving many weak components. We therefore also define the probability gap between the largest and second-largest cluster probabilities. A small gap indicates that no single regime clearly dominates.

For quantities that require robust normalization across the search domain, we use percentile-based normalization. The boundary strength at scale K is then defined as a weighted combination of high entropy and a small dominant–second probability gap. The larger weight on entropy reflects its role as the primary measure of regime mixture, while the probability-gap term provides a complementary measure of competition between the two leading regimes.

$$H_K(x) = -\frac{1}{\log K}\sum_c p_c^{(K)}(x)\log\left[p_c^{(K)}(x)+\varepsilon\right] \quad (7)$$

$$\Delta p_K(x) = p_K^{(1)}(x) - p_K^{(2)}(x) \quad (8)$$

$$N_+(z) = \mathrm{clip}\left(\frac{z-Q_{10}(z)}{Q_{90}(z)-Q_{10}(z)}, 0,1\right) \quad (9)$$

$$N_-(z) = 1 - N_+(z) \quad (10)$$

$$B_K(x) = 0.65N_+(H_K) + 0.35N_-(\Delta p_K) \quad (11)$$

### 3.4 White-space field

Novelty is represented geometrically by the distance to existing documents, but a monotonic distance reward is undesirable because it would drive the search toward unsupported regions.

We therefore define a band-pass white-space function from the nearest-document distance $d_{nn}$. Let $q_{20}$, $q_{60}$, $q_{85}$, and $q_{98}$ denote the corresponding empirical quantiles of $d_{nn}$ over the candidate domain. Very small gaps receive no novelty reward, an intermediate band receives the maximum value, and extremely remote regions are progressively rejected.

This definition implements the distinction incremental → supported novelty → unsupported extrapolation. The percentile-based formulation makes the criterion relative to the sampling density of the corpus rather than dependent on an absolute map-distance threshold.

$$W(x) = 0(d_{\mathrm{nn}} \le q_{20});\ \text{linear rise}\ q_{20} \to q_{60};\ 1\text{for}\ q_{60} \le d_{\mathrm{nn}} \le q_{85};\ \text{linear fall}\ q_{85} \to q_{98};\ 0(d_{\mathrm{nn}} \ge q_{98}) \quad (12)$$

### 3.5 Knowledge-gradient field

The continuous semantic field provides a local measure of how rapidly knowledge content changes with position. Let $J_v(x)$ be the Jacobian of the interpolated semantic field. We define the scalar gradient magnitude by the Frobenius norm and normalize it using $N_+$.

Large G identifies locations at which a small displacement on the map produces a relatively large change in semantic content. This quantity should not be interpreted as a physical gradient governed by a conservation law. Rather, it is an induced gradient of the interpolated semantic representation and is used as a diagnostic of rapid knowledge transition.

$$J_v(x) = \left[\frac{\partial v}{\partial x}\frac{\partial v}{\partial y}\right] \quad (13)$$

$$g(x) = \|J_v(x)\|_F = \sqrt{\left\|\frac{\partial v}{\partial x}\right\|_2^2 + \left\|\frac{\partial v}{\partial y}\right\|_2^2} \quad (14)$$

$$G(x) = N_+\big(g(x)\big) \quad (15)$$

### 3.6 Evidential support

A discovery candidate should not depend excessively on one or two documents. We therefore define support using three complementary quantities derived from the local interpolation neighborhood.

If $w_i(x)$ are normalized contribution weights, the effective number of contributing documents is the inverse Simpson concentration. A large $N_{eff}$ indicates broadly distributed support. Second, $d_5(x)$ denotes the distance to the fifth-nearest observed document; a small $d_5$ indicates that a candidate has several nearby sources rather than only one close neighbor. Third, $T_3(x)$ is the contribution concentration of the three highest-weight documents; a large $T_3$ indicates that the interpolation is dominated by a few documents.

These quantities are combined into the support score. Thus, a high support score favors candidates receiving evidence from many documents, having several nearby neighbors, and avoiding excessive dependence on only a few sources.

$$N_{\text{eff}}(x) = \frac{1}{\sum_i w_i(x)^2} \quad (16)$$

$$T_3(x) = \sum_{i\in\text{top3}} w_i\,(x) \quad (17)$$

$$S(x) = 0.40N_+(N_{\text{eff}}) + 0.30N_-(d_5) + 0.30N_-(T_3) \quad (18)$$

### 3.7 Model-based reliability

Support and statistical uncertainty are related but distinct. Support describes the local distribution of documents contributing to the candidate, whereas uncertainty measures how strongly a probabilistic surrogate regards the location as interpolative rather than extrapolative.

Following the previous framework, GPR provides a relative uncertainty field u(x). To assess sensitivity to covariance assumptions, two kernels are considered: the baseline anisotropic RBF model and a Matérn-3/2 alternative. Their normalized low-uncertainty scores are averaged to define R.

R is retained as a reliability diagnostic, rather than as one of the final Pareto objectives. This distinction is important because directly maximizing low uncertainty tends to favor densely sampled regions and can therefore work against the exploration of meaningful frontier regions.

$$R(x) = 0.5N_-(u_{\text{RBF}}) + 0.5N_-(u_{\text{M32}}) \quad (19)$$

## 4. Multi-Objective Search for Supported Knowledge Gaps

### 4.1 From scalar discovery scores to Pareto search

During development of the framework, candidate locations were initially ranked by a weighted discovery score combining boundary strength, white space, support, semantic gradient, reliability, and agreement between uncertainty models. Such scalarization is useful for screening, but it has an important conceptual limitation: the weights determine how different kinds of scientific opportunity are traded against one another.

There is no empirical or theoretical basis for asserting, for example, that an increase in boundary strength is scientifically equivalent to a prescribed decrease in white-space novelty. The weight vector is therefore normative rather than learned.

This is precisely the situation for which Pareto analysis is appropriate. Rather than defining one universal discovery score, Knowledge Manifold Optimization retains the competing objectives separately and searches for locations for which no objective can be improved without sacrificing at least one other objective.

$$D(x) = 0.25B + 0.20W + 0.20S + 0.15G + 0.15R + 0.05C \quad (20)$$

### 4.2 Persistent boundary objective

A knowledge boundary that appears at only one clustering resolution or only under one smoothing scheme may be an artifact of the representation. We therefore require boundary evidence to persist across three complementary views: a broad discovery-scale partition, a finer global reference partition, and a finer adaptive reference partition.

The persistent boundary objective is defined as the harmonic mean of the three boundary-strength measures. This choice is intentional because boundary persistence is an AND-like requirement: a candidate should receive a high score only when boundary evidence remains substantial in all three representations. Unlike the arithmetic mean, which permits a high value in one view to compensate for a low value in another, the harmonic mean strongly penalizes a weak component and therefore acts as a conservative aggregation operator. This is consistent with the purpose of $P_{\text{boundary}}$, which is to identify boundaries that persist across clustering resolutions and smoothing scales rather than those produced by a single representation. No matching of cluster labels across resolutions is required; only the scalar strength of local regime competition is compared.

$$P_{\text{boundary}}(x) = \frac{3}{B_{5,g}(x)^{-1} + B_{8,g}(x)^{-1} + B_{8,a}(x)^{-1}} \quad (21)$$

### 4.3 Final discovery-objective vector

The final optimization problem retains four objectives with distinct scientific interpretations. Reliability R is not included because it partially overlaps with support and, if maximized, can over-favor densely populated regions. Likewise, uncertainty-model consistency is treated as a sensitivity diagnostic rather than as a scientific-discovery objective.

$$\text{maximize over } x \in \Omega^*: F(x) = \left[P_{\text{boundary}}(x), W(x), G(x), S(x)\right] \quad (22)$$

The objectives answer four different questions: Does the location persistently bridge knowledge regimes? Is it novel without being an extreme extrapolation? Does knowledge change rapidly around the location? Is the inferred content supported by a sufficiently broad neighborhood?

### 4.4 Pareto dominance

For two candidate locations $x_a$ and $x_b$, $x_a$ dominates $x_b$ if it is no worse in all objectives and strictly better in at least one. The Pareto set therefore contains locations for which no admissible alternative dominates them.

A key interpretation is that the Pareto set is not expected to identify one universally optimal discovery location. Instead, different Pareto points represent different discovery archetypes, in the same sense that different Pareto-optimal engineering designs represent different compromises among conflicting design objectives. This interpretation provides the direct connection between Multi-Objective Design Exploration and Knowledge Manifold Optimization.

$$F_j(x_a) \geq F_j(x_b) \text{for all } j, F_j(x_a) > F_j(x_b) \text{for at least one } j \quad (23)$$

$$\mathcal{P} = \{x \in \Omega^* : \text{there exists no } y \in \Omega^* \text{ such that } y \text{ dominates } x\} \quad (24)$$

### 4.5 Practical Pareto front

Pareto dominance alone can preserve mathematically non-dominated points that are scientifically uninteresting. In particular, because white-space novelty is deliberately band-pass filtered, candidates with $W(x)=0$ correspond either to regions that are too close to existing documents or to regions that are too remote to be regarded as supported knowledge gaps.

We therefore define the practical Pareto front as the subset of non-dominated points satisfying $W(x)>0$.

Candidate points located on or very near the outer boundary of the fixed chart are not excluded from Pareto analysis. However, they are treated cautiously in subsequent scientific interpretation because their geometry may be influenced by the bounded chart and the gauge-fixing anchor configuration. Accordingly, an outer-boundary candidate may remain mathematically non-dominated while receiving a more conservative interpretation as a scientific discovery opportunity.

$$\mathcal{P}_{\text{practical}} = \{x \in \mathcal{P} : W(x) > 0\} \quad (25)$$

### 4.6 Exact search on a discretized Knowledge Manifold

For the present study, the four discovery objectives are evaluated on a regular grid over the fixed two-dimensional chart. Because the number of candidate locations is modest, the Pareto front can be obtained by exact pairwise dominance testing. No evolutionary multi-objective optimizer is therefore required for the present analysis.

This choice avoids an unnecessary additional source of stochastic variability and ensures that the extracted discrete Pareto front is complete for the sampled grid. For a future continuous formulation, the map coordinates could instead be treated directly as continuous decision variables and searched using algorithms such as NSGA-II [34] or other multi-objective global optimizers.

### 4.7 From Pareto solutions to discovery candidates

The Pareto front is interpreted as a structured set rather than a ranked list. Nearby Pareto points may represent the same broad scientific opportunity, whereas separated regions of the front may correspond to qualitatively different mechanisms of discovery.

Accordingly, interpretation proceeds in two stages. First, all practical Pareto points are retained and examined in objective space and on the Knowledge Manifold. Second, groups of semantically related points are interpreted as discovery archetypes, from which representative locations can be selected for detailed evidence extraction.

For a representative location x*, the Knowledge Manifold exports the semantic terms and their evidential provenance, neighboring document contributions, cluster-regime mixture, effective support, gradient magnitude, and uncertainty measures. Only after this evidence package has been generated is the candidate passed to the language model for verbalization.

$$\text{Pareto location} \rightarrow \text{evidence package} \rightarrow \text{candidate scientific hypothesis} \quad (26)$$

## 5. Aerodynamic Literature Case Study

### 5.1 Corpus construction

The proposed Knowledge Manifold Optimization framework was applied to a corpus of research papers in aerodynamic optimization. The purpose of this case study was not to construct an exhaustive bibliometric database of aerodynamics, but to examine whether scientifically interpretable discovery opportunities could be identified from a realistically sized, domain-focused literature corpus.

The corpus was assembled in three stages. First, LeapSpace was used to recommend 150 papers on aerodynamic optimization published during the preceding ten years. Second, among these recommended papers, full-text PDF files that could be downloaded through Scopus were collected. Finally, review articles were excluded because the present framework treats each document as an individual observation of research activity, whereas review articles intentionally aggregate broad areas of prior work and could therefore distort local semantic density and interpolation.

After these screening steps, the final corpus contained N = 99 research papers.

The resulting corpus should be regarded as a domain-focused discovery corpus rather than a systematic-review sample. Its composition is affected both by the recommendation procedure used by LeapSpace and by full-text availability through Scopus. Accordingly, all novelty claims made in this paper are explicitly relative to these 99 documents.

This distinction becomes important when interpreting the final Supported Knowledge Gaps: they represent gaps in the analyzed aerodynamic-optimization corpus and not proof that the corresponding research topics have never appeared elsewhere in the scientific literature.

$$\text{150 LeapSpace recommendations} \rightarrow \text{downloadable PDFs} \rightarrow \text{removal of review papers} \rightarrow \text{99 research papers} \quad (27)$$

### 5.2 Document representation and fixed Knowledge Manifold

PDF text was extracted without OCR and represented by within-word character n-gram TF–IDF features with n = 4–7, sublinear term frequency, smoothed inverse document frequency, and a maximum vocabulary size of 250,000 features. The same raw representation was normalized using both $\ell_2$ and $\ell_1$ norms, following the Knowledge Manifold formulation [28,29].

The $\ell_2$ representation was used as the principal semantic geometry for map construction and knowledge-gradient evaluation. The $\ell_1$ representation provided a complementary feature-allocation view and was subsequently used during evidence interpretation.

All 99 papers were placed on one fixed bounded two-dimensional Knowledge Manifold, $\Omega = [-1,1]^2$. Nine automatically selected anchor documents fixed the coordinate gauge, consisting of eight perimeter anchors and one central anchor. The remaining documents were positioned through constrained stress minimization initialized by multidimensional scaling. Once obtained, the coordinates were frozen throughout the discovery analysis.

This is important: clustering resolution, SPH smoothing, GPR kernel choice, and discovery objectives were varied during sensitivity analyses, but the underlying document coordinates were not recomputed. Consequently, all comparisons refer to the same knowledge-space locations.

## 5.3 Multi-resolution structure of the aerodynamic knowledge space

Cluster structures with K = 3 through K = 9 were examined on the same fixed map. The purpose was not to determine a statistically unique number of clusters, but to identify resolutions that were useful for different levels of discovery analysis.

A five-cluster representation was selected as the Discovery Level, because it provided a compact division into broad research regimes while retaining a mixed region that could act as a bridge between established fields. An eight-cluster representation was selected as the Reference Level, because it resolved several specialized aerodynamic and optimization subfields without the excessive fragmentation observed at K = 9.

Cluster identifiers are local to each resolution and have no assumed semantic equivalence across different K. Cross-scale comparisons are therefore based on document membership and continuous boundary fields rather than numerical cluster labels.

The K = 9 representation produced additional splitting dominated in part by near-duplicate title pairs and was therefore judged too fragmented for the present purpose. Conversely, K = 5 provided a useful wide-angle view, while K = 8 provided a finer semantic reference. These choices were based on semantic coherence, persistence, and interpretability rather than on the two-dimensional silhouette coefficient alone. The selected Discovery and Reference resolutions are compared in Fig. 2.

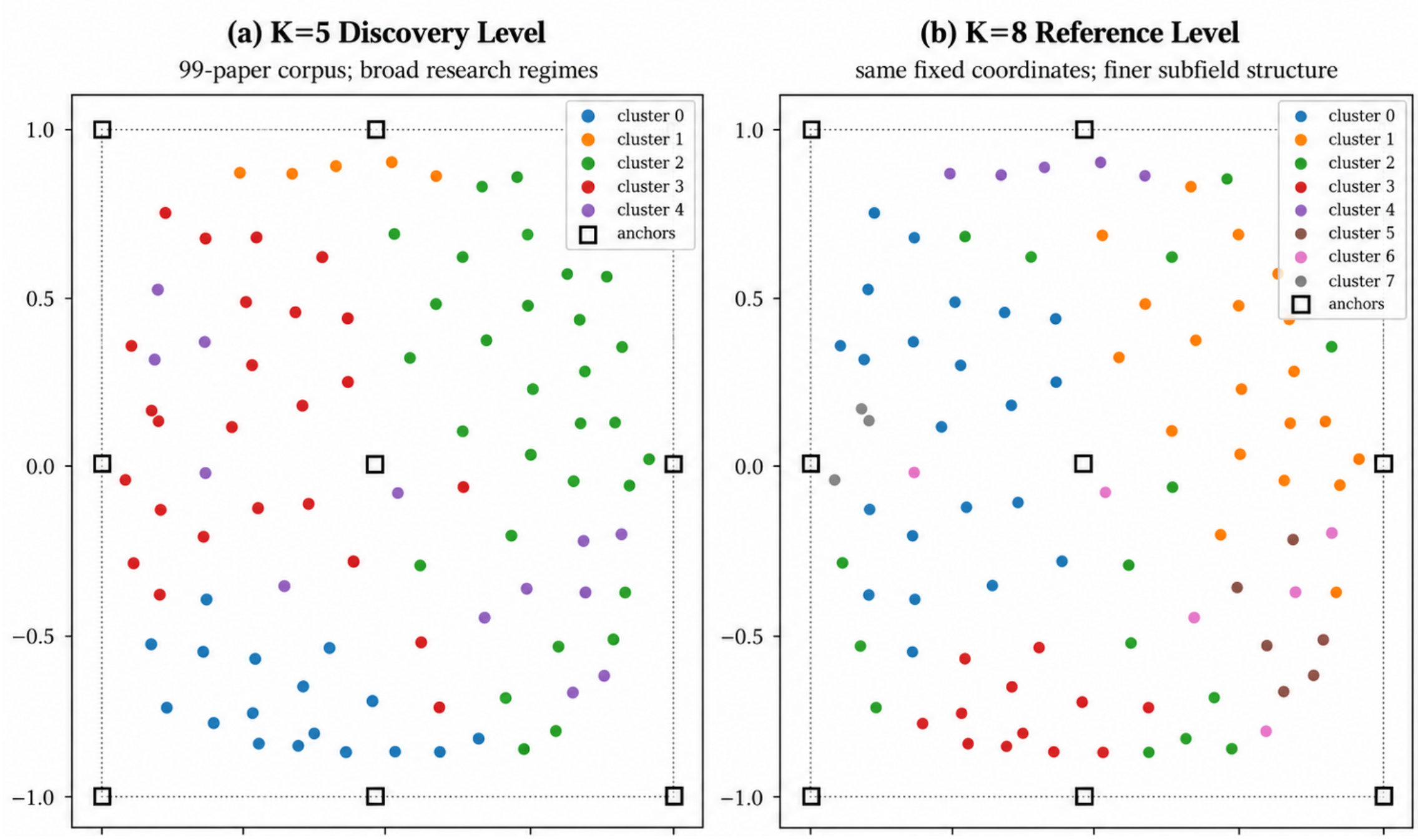

Figure 2. Knowledge Manifold of the 99-paper aerodynamic-optimization corpus at two clustering resolutions. The K=5 map is used as the broad Discovery Level and K=8 as the finer Reference Level. The document coordinates are identical in both panels; only the clustering resolution changes. Open squares denote gauge-fixing anchors.

Table 1. Multi-resolution cluster interpretations of the aerodynamic-optimization Knowledge Manifold.

| Resolution | Cluster | Main interpretation |
|---|---|---|
| K = 5 | C0 | topology optimization / FSI foundations |
| | C1 | persistent frontier group |
| | C2 | applied aerodynamics, high-lift, drag and flow |
| | C3 | AI/ML, surrogate and multifidelity optimization |
| | C4 | mixed bridge / conventional shape optimization and applied CFD |
| K = 8 | C0, n=27 | core aerodynamic shape optimization methods |
| | C1, n=24 | high-lift, wings, flaps and flow control |
| | C2, n=16 | topology creation and advanced design methods |
| | C3, n=11 | fluid–structure-interaction topology |
| | C4, n=6 | persistent frontier group |
| | C5, n=6 | vehicle aerodynamics and drag reduction |
| | C6, n=5 | integrated / multiphysics / multi-objective optimization |
| | C7, n=4 | high-dimensional and data-efficient AI |

## 5.4 Global and adaptive interpolation

Two SPH-based representations of local regime mixture were evaluated. The global field used one smoothing scale over the full map and was intended to capture broad research continuity. The adaptive field used a local scale based on the eight nearest documents and was intended to test whether apparent boundary regions remained mixed when the neighborhood was localized.

This distinction is important because a candidate may appear interdisciplinary under broad smoothing while collapsing into one dominant regime under local inspection. Such points are less convincing as persistent discovery boundaries.

At the detailed K = 8 resolution, the adaptive dominant-cluster counts over the search grid were 166, 130, 35, 46, 26, 29, 0, and 9 for C0–C7, respectively. The absence of C6 as a locally dominant cluster does not imply absence of the corresponding integrated/multiphysics regime. C6 contributes continuously to several bridge regions but is generally accompanied by a stronger neighboring regime.

## 5.5 Evaluation grid and discovery-field parameters

The discovery fields were evaluated on a regular 21 × 21 grid covering $[-1,1]^2$, giving 441 candidate locations with a grid spacing of 0.1.

Robust normalization used the 10th and 90th percentiles as defined in Eqs. (9)–(10). For the white-space objective, the nearest-document-distance quantiles were Q20 = 0.0517, Q60 = 0.1000, Q85 = 0.1376, and Q98 = 0.2000. Thus, candidate locations with nearest-document distances between approximately 0.10 and 0.138 map units received the maximum white-space score, while very close and extremely remote regions were progressively suppressed.

GPR uncertainty was evaluated using both the baseline anisotropic RBF kernel and a Matérn-3/2 kernel. Neither uncertainty field was used to generate the semantic content of a candidate; both served as reliability diagnostics for the interpolated Knowledge Manifold.

Importantly, no language model was used in the calculation of any discovery objective or in Pareto-front extraction. The resulting four discovery fields are shown in Fig. 3.

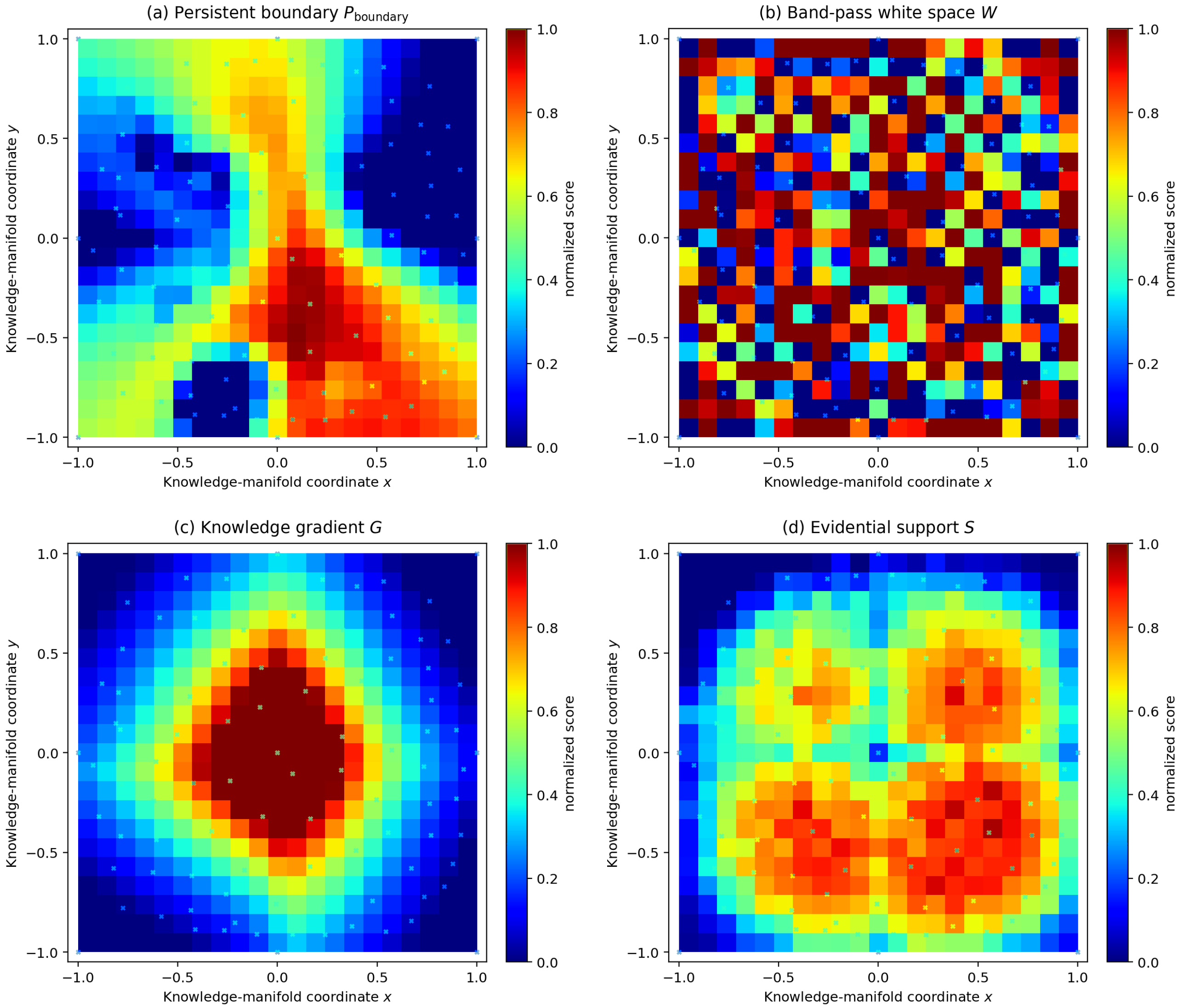

Figure 3. Four normalized discovery fields used in Knowledge Manifold Optimization: persistent boundary $P_{boundary}$, band-pass white space W, knowledge gradient G, and evidential support S. Crosses mark the 99 document positions. The white-space field rewards intermediate separation from existing documents rather than maximum emptiness.

$$Q_{20} = 0.0517, Q_{60} = 0.1000, Q_{85} = 0.1376, Q_{98} = 0.2000 \quad (28)$$

# 6. Pareto-Optimal Supported Knowledge Gaps

## 6.1 Reduction from scalar ranking to a practical Pareto front

The analysis was initially conducted using the six quantities introduced during exploratory development: boundary strength, white space, support, semantic gradient, reliability, and uncertainty-model consistency.

When all six were treated independently as objectives, however, 139 of 441 grid points were non-dominated, corresponding to approximately 31.5% of the search space. Such a large Pareto set provides insufficient discrimination for practical discovery.

The objective system was therefore reduced on conceptual grounds rather than by fitting the observed results. Reliability and uncertainty-model consistency were retained as validation quantities, while the four quantities directly describing scientific-discovery opportunity were retained as Pareto objectives: [$P_{boundary}$, W, G, S]. This reduced the non-dominated set to 19 points. Three of these had W = 0, indicating locations outside the intended band of supported novelty. Excluding these yielded a practical Pareto front of 16 locations.

$$441 \rightarrow 139 \rightarrow 19 \rightarrow 16 \quad (29)$$

## 6.2 Practical Pareto front

The sixteen practical Pareto solutions are listed in Table 2. No single point dominates the front in all four objectives. Reliability R is included for interpretation but is not an optimization objective.

Table 2. Practical Pareto solutions identified by Knowledge Manifold Optimization.

| (x, y) | $P_{boundary}$ | W | G | S | R |
|---|---|---|---|---|---|
| (-0.4, -0.3) | 0.434 | 1.000 | 0.710 | 0.909 | 0.759 |
| (-0.2, -0.3) | 0.671 | 1.000 | 1.000 | 0.766 | 0.781 |
| (0.0, -0.5) | 0.958 | 0.357 | 0.914 | 0.792 | 0.647 |
| (0.1, -0.5) | 0.994 | 0.888 | 0.897 | 0.851 | 0.685 |
| (0.1, -0.4) | 0.984 | 0.888 | 1.000 | 0.748 | 0.702 |
| (0.1, -0.2) | 0.973 | 0.992 | 1.000 | 0.721 | 0.712 |
| (0.2, -0.4) | 0.976 | 0.561 | 0.920 | 0.761 | 0.729 |
| (0.2, -0.2) | 0.969 | 1.000 | 1.000 | 0.699 | 0.711 |
| (0.2, -0.1) | 0.903 | 1.000 | 1.000 | 0.754 | 0.895 |
| (0.3, -0.4) | 0.942 | 1.000 | 0.762 | 0.903 | 0.767 |
| (0.3, -0.3) | 0.925 | 0.968 | 0.877 | 0.867 | 0.769 |
| (0.3, -0.2) | 0.884 | 1.000 | 0.994 | 0.811 | 0.740 |
| (0.4, -0.6) | 0.922 | 1.000 | 0.449 | 0.905 | 0.767 |
| (0.4, -0.3) | 0.853 | 1.000 | 0.698 | 0.943 | 0.797 |
| (0.4, -0.2) | 0.761 | 1.000 | 0.784 | 0.864 | 0.808 |
| (0.6, -0.2) | 0.551 | 1.000 | 0.464 | 0.949 | 0.905 |

For example, (0.2,−0.1) combines maximum white-space and gradient values with high reliability, whereas (0.1,−0.5) exhibits exceptionally persistent cross-scale boundary structure together with stronger support. Point (0.6,−0.2) has very high support and reliability but substantially lower boundary and gradient values.

The point (0.3,−0.3) is closest to the ideal corner under an equal-distance objective-space criterion and may therefore be regarded as a balanced compromise. It is not, however, interpreted as a universally best discovery point, because such an interpretation would reintroduce an implicit scalar preference among the objectives. Their distribution on the Knowledge Manifold is shown in Fig. 4.

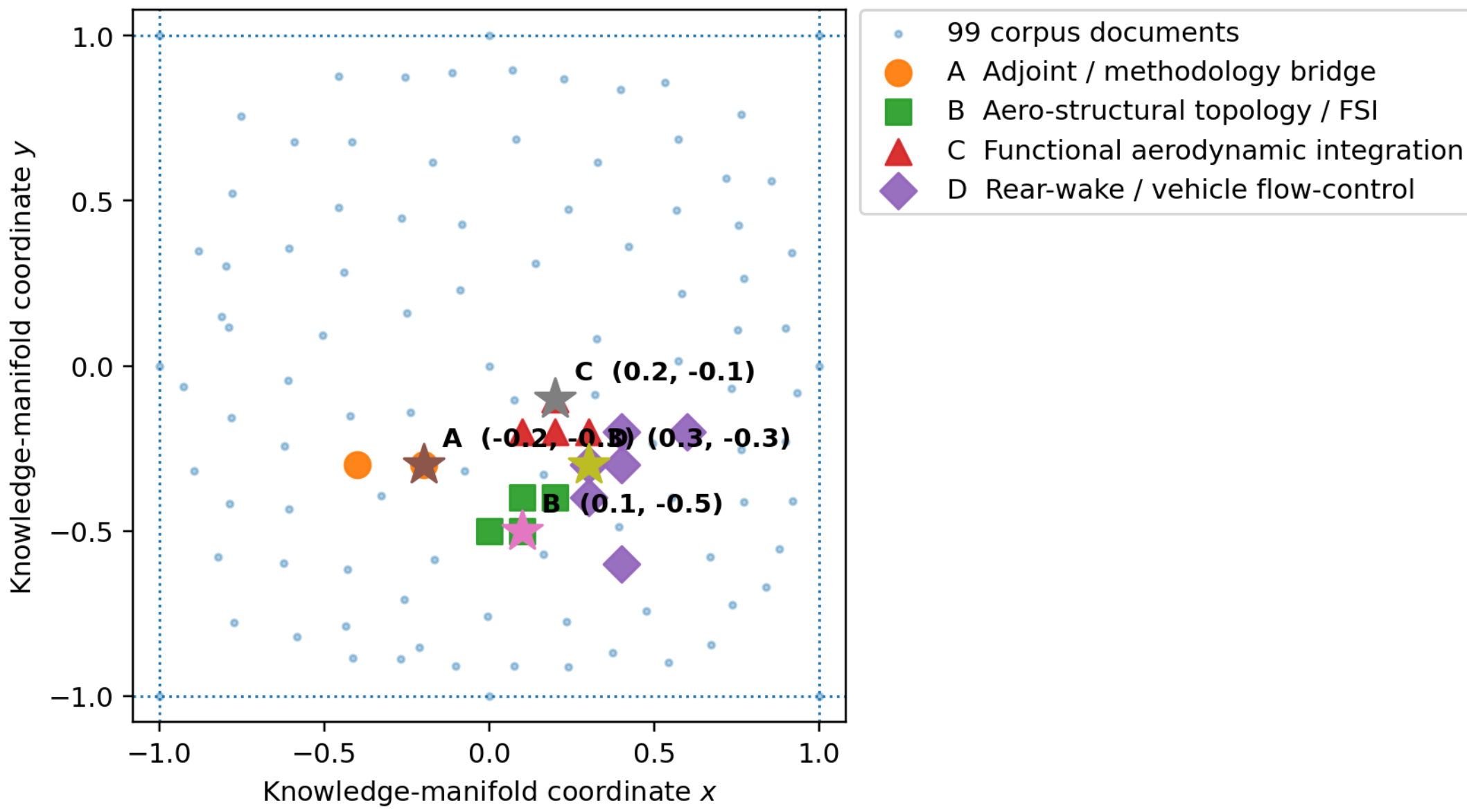

Figure 4. Sixteen practical Pareto solutions projected onto the fixed Knowledge Manifold. Marker shapes indicate the four interpretive discovery archetypes A-D, and stars denote representative locations. The archetypes are interpretive groupings of the Pareto set rather than outputs of an additional clustering algorithm.

### 6.3 Discovery archetypes

The practical Pareto solutions form several spatially and semantically related groups. Based jointly on their locations on the Knowledge Manifold, contributing documents, local semantic evidence, and multi-resolution cluster-mixture profiles, four discovery archetypes were identified as an interpretive summary of the non-dominated set.

Table 3. Interpretive discovery archetypes of the practical Pareto front.

| Archetype | Pareto region | Representative | Interpretation |
|---|---|---|---|
| A | (-0.4,-0.3), (-0.2,-0.3) | (-0.2,-0.3) | adjoint / methodology bridge |
| B | (0,-0.5) to (0.2,-0.4) | (0.1,-0.5) | aero-structural topology / FSI |
| C | (0.1,-0.2) to (0.3,-0.2) | (0.2,-0.1) | functional aerodynamic devices / attachment–vortex integration |
| D | (0.3,-0.4) to (0.6,-0.2) | (0.3,-0.3) | rear-wake / vehicle flow-control integration |

The four archetypes should not be interpreted as algorithmically determined clusters of the Pareto set. Rather, they constitute a post hoc interpretive typology used to summarize distinct modes of discovery represented by the non-dominated solutions. As an independent geometric reference, four-cluster Ward hierarchical clustering was applied separately in the four-dimensional objective space [$P_{boundary}$, W, G, S] and in the two-dimensional manifold-coordinate space. The resulting partitions showed partial agreement with the four interpretive archetypes, indicating that the typology is supported by geometric organization but is not reducible to Euclidean proximity alone. The final archetype interpretation additionally incorporates local semantic evidence, contributing-document provenance, and multi-resolution cluster-mixture structure. Accordingly, the archetype labels are used only for scientific interpretation and do not affect Pareto optimality, candidate selection, or any numerical result.

The result illustrates a central motivation for Pareto analysis. The discovery landscape does not contain one type of optimum. Instead, different non-dominated locations represent distinct mechanisms by which new knowledge may emerge. Their trade-offs in objective space are shown in Fig. 5.

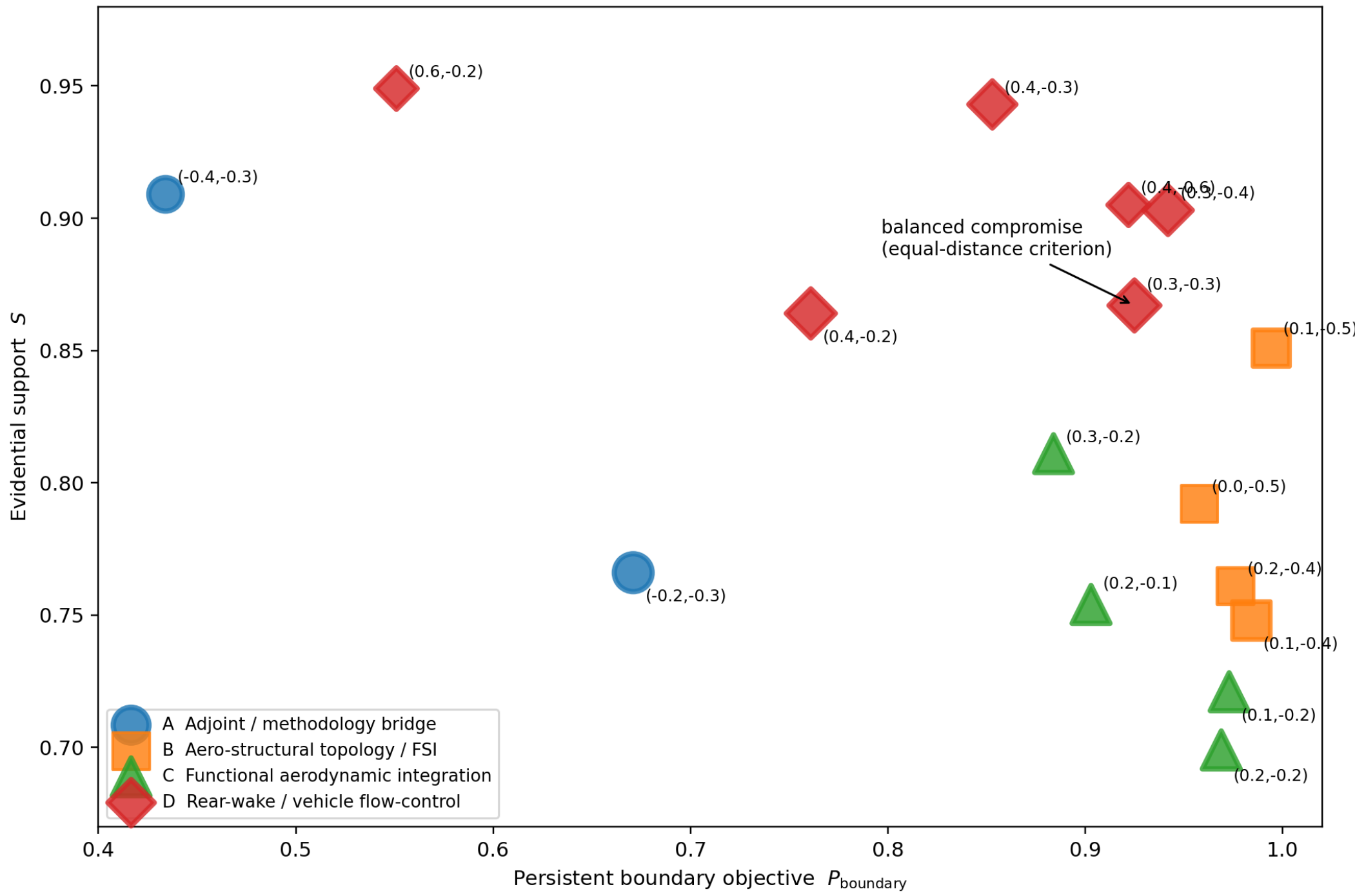

Figure 5. Objective-space interpretation of the practical Pareto front. The horizontal axis is persistent boundary $P_{boundary}$, the vertical axis is evidential support S, bubble size represents knowledge-gradient objective G, and outline thickness represents white-space objective W. The highlighted point (0.3,-0.3) is a balanced compromise under an equal-distance criterion; no unique best solution is implied.

## 6.4 Effect of cross-scale validation

Cross-scale analysis provides additional discrimination among candidates that appeared strong under the earlier weighted score.

At x = (0.2,−0.1), the global K = 8 mixture was approximately (C1,C0,C2,C6) = (0.302,0.251,0.176,0.087), while the adaptive representation gave (C1,C2,C0,C6) = (0.307,0.252,0.240,0.161). The local field therefore remained genuinely mixed after adaptive localization.

Similarly, at x = (0.1,−0.5), the global mixture was approximately (C0,C3,C2,C1) = (0.250,0.209,0.202,0.164), whereas the adaptive field gave (C2,C3,C0,C6) = (0.305,0.284,0.247,0.107). Again, several regimes remained competitive.

By contrast, two candidates that had scored highly in the initial scalar search exhibited considerably stronger local collapse. At (−0.2,−0.3), the adaptive C0 probability increased to approximately 0.646, with a dominant–second gap of 0.453. At (−0.2,0.3), the adaptive C0 probability reached approximately 0.775 and the gap increased to 0.654. The latter point was also dominated under the final four-objective Pareto formulation.

This comparison demonstrates why neither a single clustering resolution nor a single weighted discovery score is sufficient. Persistent cross-scale mixing provides a stronger indication that a candidate truly lies between knowledge regimes. The multi-resolution probability profiles are summarized in Fig. 6.

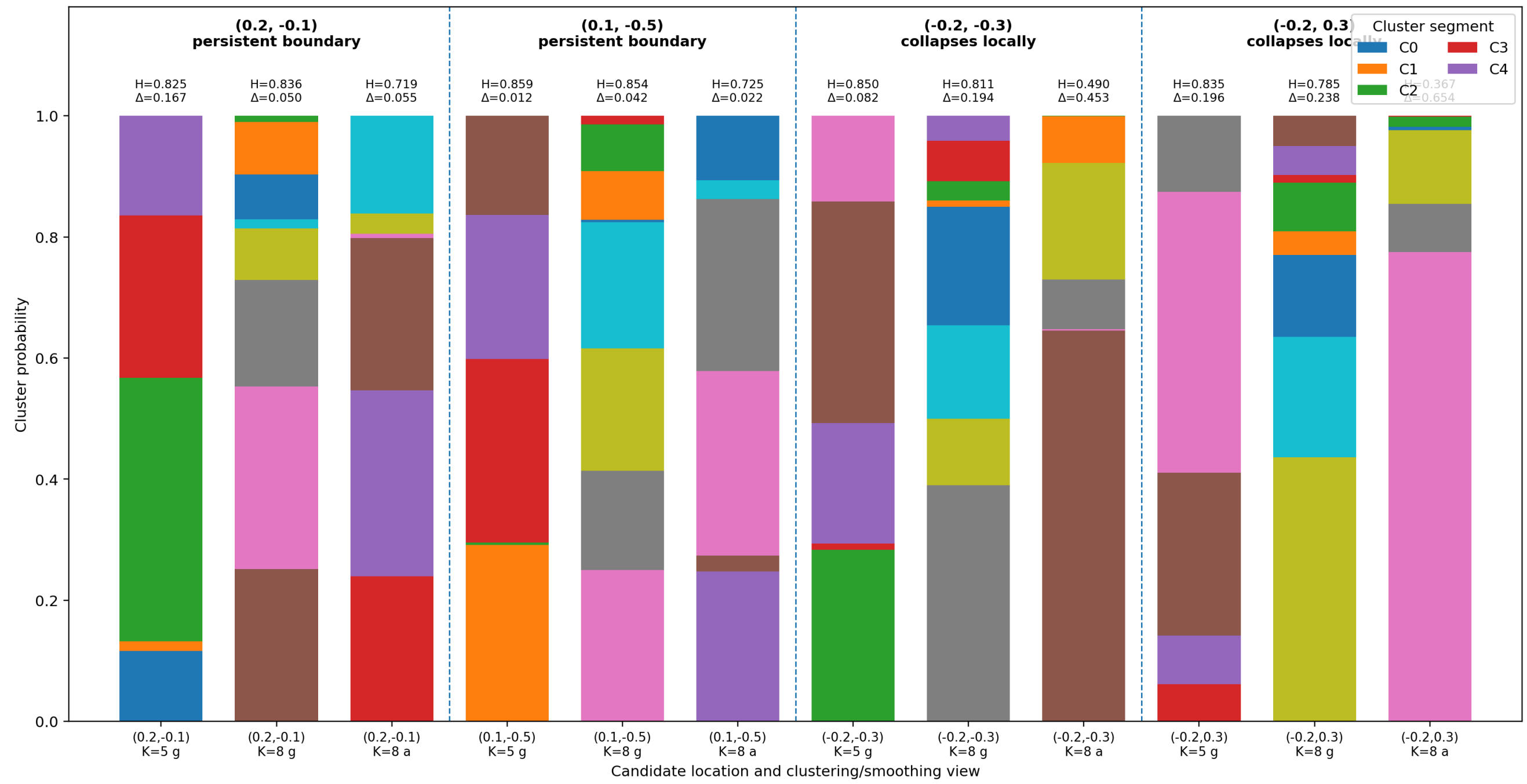

Figure 6. Multi-resolution boundary validation for representative candidates. Stacked bars show cluster-probability mixtures under K=5 global, K=8 global, and K=8 adaptive views; H denotes normalized mixture entropy and Delta the dominant-second probability gap. Persistent candidates retain mixed local regimes, whereas collapse candidates become dominated by a single regime under adaptive localization.

# 7. Evidence-Constrained Interpretation of Discovery Candidates

## 7.1 Evidence package

Pareto optimization determines where potentially interesting knowledge gaps occur, but coordinates alone do not explain what scientific content such a location represents.

For each selected Pareto point $x^*$, an evidence package is therefore exported directly from the Knowledge Manifold. The package contains the local semantic vocabulary, document-level contributions, document frequencies of relevant terms, multi-resolution regime probabilities, effective document support, and GPR uncertainty.

Two semantic lenses are used. Terms that are prominent in both the $\ell_2$ and $\ell_1$ representations are treated as the strongest local thematic evidence. Terms appearing primarily in the $\ell_1$ representation are interpreted more cautiously because this representation is more sensitive to concentrated or relatively rare feature allocation.

In particular, an $\ell_1$-specific term with document frequency $df \geq 4$ may be used as an emergent shared topic, whereas a term with $df \leq 3$ is admitted only as a limited-source hypothesis and its contributing source documents must be stated explicitly. This protocol is designed to prevent a rare token from being promoted into an established regional theme.

## 7.2 Evidence-constrained verbalization

In the controlled verbalization task reported here, the language model receives the exported evidence package together with a fixed researcher-supplied protocol. The language model does not choose the candidate location and is not asked to retrieve additional literature during this stage.

The protocol imposes explicit evidential constraints. The L2 theme lens and the contributing documents define the consensus topic; terms appearing in both the L2 and L1 lenses receive the greatest weight. L1-only terms with document frequency $df \geq 4$ may be treated as emergent shared topics, whereas terms with $df \leq 3$ may be introduced only as limited-source hypotheses with their source documents stated explicitly. The model is instructed not to introduce outside scientific facts, to use every supplied term or explicitly account for its exclusion, and to end with an Evidence used list.

The principal output is a 200–280-word virtual research abstract describing the candidate topic, methodological direction, and expected contribution, followed by an exactly 45-word summary. The abstract must include a caveat reflecting relative uncertainty and the effective number of contributing documents. For the two examples reported here, the generated output was then audited against the evidence package for word-count compliance, use and provenance of supplied terms, treatment of low-frequency terms, unsupported technical claims, novelty wording, and explicit uncertainty reporting. The complete prompt and audit checklist are given in Appendix A.

$$\text{Pareto optimization} \rightarrow \text{evidence extraction} \rightarrow \text{constrained verbalization} \quad (30)$$

These constraints do not constitute a formal guarantee that an LLM can never hallucinate. Their purpose is instead to make the generation evidence-constrained and auditable: every substantive statement in the virtual abstract can be compared with a finite, machine-exported evidence package. The resulting text is therefore a corpus-grounded hypothesis proposal rather than evidence of external scientific novelty or validity.

### 7.3 Representative candidate I: functional aerodynamic integration

The Pareto point $x = (0.2, -0.1)$ was selected as a representative of Archetype C.

Its local semantic field contains the terms rear, airflow, attachment, enhances, angles, vortex, and circulation, with additional evidence associated with biplane configurations. The most strongly contributing documents were docs 73, 63, 58, 89, and 8.

The location simultaneously lies near the interface of the high-lift / flow-control, advanced-design, and core optimization regimes. Its normalized gradient reaches the maximum value in the search domain, $G = 1.0$, while its white-space value is also $W = 1.0$. The RBF relative uncertainty was approximately $u_{RBF} = 0.222$, and the effective support was approximately 74 documents.

Taken together, the evidence suggests a research opportunity in which advanced aerodynamic optimization is connected to functional flow-modifying devices characterized by attachment, circulation, vortex behavior, and rear-element geometry.

Importantly, a rare term such as Trefftz, associated with only one contributing document in the local evidence, is not treated as an established regional concept. It can only be introduced, if at all, as a source-specific hypothesis. This example is therefore characterized primarily by high semantic transition and cross-regime functional integration.

The complete generated 245-word virtual abstract, 45-word summary, and evidence-accounting list for this candidate are reproduced in Appendix B.1.

### 7.4 Representative candidate II: aero-structural topology and FSI

The point $x = (0.1, -0.5)$ was selected as a representative of Archetype B.

Its dominant semantic evidence includes material, topology, compliance, stiffness, structural, and penalization, with nearby contributions associated with fluid–structure interaction. The most influential neighboring documents included docs 1, 8, 48, 52, and 9.

Unlike the previous example, this candidate is characterized less by a single sharp aerodynamic phenomenon than by the integration of several methodological and physical descriptions. Its cross-scale boundary persistence is extremely high, $P_{boundary} = 0.994$, while $W = 0.888$, $G = 0.897$, and $S = 0.851$. Its effective support is approximately 72 documents, and the baseline GPR uncertainty is approximately 0.23.

The corresponding Supported Knowledge Gap can therefore be interpreted as an opportunity for aero-structural design in which topology-oriented material representation, compliance or stiffness constraints, aerodynamic optimization, and fluid–structure interaction are considered within a common design framework. Relative to the first candidate, this location represents a more methodologically coherent form of integration.

The complete generated 245-word virtual abstract, 45-word summary, and evidence-accounting list for this candidate are reproduced in Appendix B.2.

### 7.5 Different Pareto points imply different discovery mechanisms

The two representative examples illustrate why the Pareto front should not be reduced to one best candidate.

Point (0.2,−0.1) is attractive because of its strong semantic gradient and integration of functional aerodynamic phenomena with optimization methods. Point (0.1,−0.5), in contrast, exhibits stronger cross-scale boundary persistence and a coherent connection between structural topology and aerodynamic/FSI design.

The former may therefore be described as a phenomenon-integration discovery, whereas the latter is closer to a method-and-multiphysics integration discovery. Both satisfy the operational concept of a Supported Knowledge Gap, but for different reasons. This multiplicity is analogous to multi-objective engineering design: different Pareto-optimal designs encode different trade-offs, and the set itself provides more information than any single preferred solution. The complete numerical-to-verbalization workflow for these representative examples is summarized in Fig. 7.

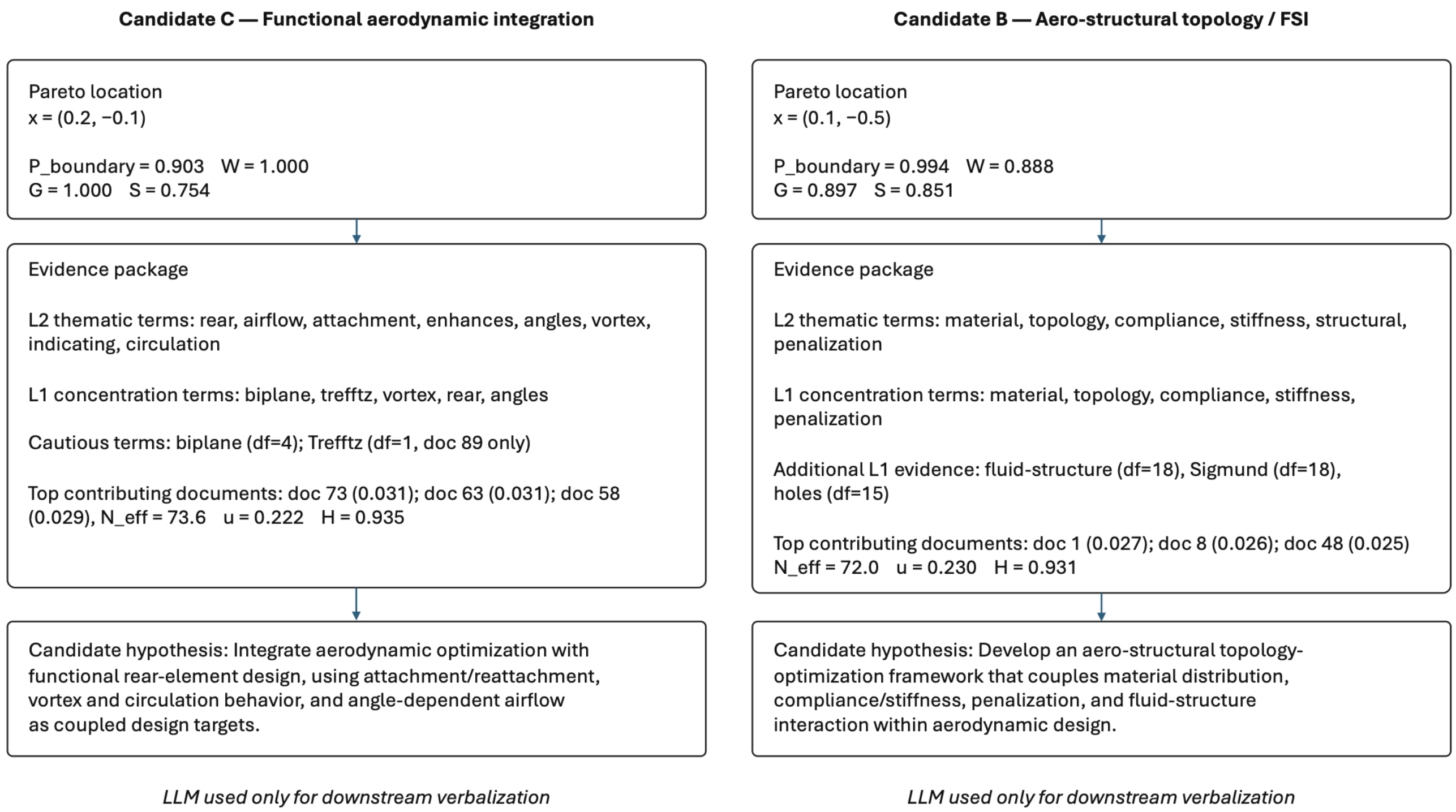

Figure 7. From Pareto-optimal locations to evidence-constrained research hypotheses for two representative candidates. The numerical search identifies the location, the Knowledge Manifold exports the evidence package, and the language model is used only downstream to verbalize that evidence into a candidate hypothesis.

## 8. Robustness and Sensitivity Analysis

### 8.1 Fidelity of the fixed Knowledge Manifold

Before interpreting discovery locations, the fidelity of the underlying two-dimensional map was examined. The final map gave a normalized stress of approximately 0.1505. The correspondence between original semantic distances and two-dimensional distances was $\rho_{\text{Spearman}} = 0.613$ and $r_{\text{Pearson}} = 0.584$.

Trustworthiness values for neighborhood sizes k = 5, 7, and 10 were approximately 0.770, 0.767, and 0.789, while the corresponding continuity values were 0.858, 0.856, and 0.855. These values indicate that the fixed two-dimensional representation retains useful global and mesoscale structure but should not be interpreted as an exact preservation of every local nearest-neighbor relation.

Accordingly, the discovery analysis relies on continuous fields, multi-resolution regime mixtures, and sensitivity checks rather than on individual two-dimensional nearest-neighbor relationships alone.

The two-dimensional chart is adopted here for interpretability and exploratory optimization rather than as a claim about intrinsic dimensionality. The KMO formulation is not inherently restricted to two dimensions and could, in principle, be extended to higher-dimensional knowledge representations, with continuous multi-objective algorithms such as NSGA-II used in place of the present two-dimensional grid search; such an extension remains to be validated.

### 8.2 Sensitivity to map initialization

Alternative random seeds produced very similar stress values, 0.1505–0.1508, indicating that overall distance preservation was stable.

Cluster agreement was less invariant, with adjusted Rand indices ranging approximately from 0.432 to 0.801. This difference is important. It shows that visually similar low-dimensional geometries do not guarantee identical discrete cluster labels.

For this reason, Knowledge Manifold Optimization does not treat a cluster identifier as a fundamental physical quantity. Continuous mixture probabilities, entropy, semantic gradients, and cross-resolution persistence are used instead. The result also supports the use of a fixed reference map throughout the present study.

### 8.3 Sensitivity to map gauge and anchoring

The anchor structure was found to matter more strongly than the random initialization. In particular, changing from a reduced four-anchor configuration to the prescribed eight perimeter anchors plus central anchor altered both map geometry and subsequent clustering sufficiently to affect local interpretation.

The fixed gauge should therefore be regarded as part of the definition of the particular Knowledge Manifold used in this study rather than as an inconsequential visualization choice. This is also why points close to the outer chart boundary require particular caution: their geometry can be influenced by the bounded chart and anchor constraints.

### 8.4 Global versus adaptive interpolation

The global and adaptive SPH models exhibit a useful trade-off. For the global field, leave-one-out cosine agreement was approximately $0.461 \pm 0.052$, with a mean cluster entropy of approximately 0.912. The corresponding adaptive field gave $0.452 \pm 0.081$ and a lower entropy of approximately 0.567.

The global field therefore produces a smoother and more strongly mixed representation suitable for broad discovery exploration, whereas adaptive interpolation resolves local specialization more sharply.

This distinction is precisely why both are retained in the persistent boundary criterion. A location that appears mixed only under global smoothing may simply reflect excessive averaging; a location that remains mixed under adaptive smoothing provides stronger evidence of a genuine local knowledge boundary.

### 8.5 GPR kernel sensitivity

The baseline uncertainty field used an anisotropic RBF kernel with a white-noise component. A Matérn-3/2 model was fitted independently as a sensitivity model.

The two uncertainty maps were strongly, but not perfectly, correlated, with $r = 0.851$. The Matérn model achieved a higher log marginal likelihood, −853.76, than the baseline RBF model, −880.9, and also showed improved latent-space leave-one-out error characteristics.

Nevertheless, the RBF formulation was retained as the reference model because it formed part of the original Knowledge Manifold analysis, while the Matérn result was used as an independent robustness check rather than as a post-hoc replacement. The high cross-model correlation indicates that major low- and high-uncertainty regions are not artifacts of one covariance choice.

### 8.6 Sensitivity to clustering resolution

The broad conclusions were also examined over $K = 3–9$. At low K, distinct specialized research regimes were merged, reducing the ability to identify interdisciplinary boundaries. At high K, particularly $K = 9$, small clusters became increasingly sensitive to near-duplicate documents and local fragmentation.

The five-cluster representation therefore served as a useful broad-scale discovery map, whereas the eight-cluster representation provided a more detailed reference map. The fact that some high-scoring $K = 5$ candidates collapsed toward a single regime under adaptive K = 8 interpolation illustrates why apparent knowledge boundaries should not be inferred from one clustering level alone.

### 8.7 Sensitivity to the objective formulation

The transition from six independent objectives to four discovery objectives also provides a sensitivity test of the formulation. With six objectives, 139 of 441 locations were non-dominated. Such a large front is mathematically valid but operationally weak because almost one-third of the grid remains Pareto-optimal.

After separating discovery attributes ($P_{boundary}$, W, G, S) from reliability diagnostics (R, C), the non-dominated set contracted to 19 points and the practical set to 16.

This reduction is not simply a numerical convenience. It reflects a conceptual distinction: uncertainty and model agreement indicate how confidently a candidate can be interpreted, whereas boundary structure, white space, semantic transition, and support describe why it may be scientifically interesting.

The aggregation operator used in $P_{boundary}$ was also tested explicitly. Replacing the harmonic mean with the arithmetic mean changed the full four-objective Pareto set from 19 to 20 points and the practical front from 16 to 17 points, so the overall structure of the result was stable. The difference is nevertheless conceptually meaningful: the arithmetic mean is more compensatory and therefore allows some candidates with weak boundary evidence in one representation to remain competitive. For example, at x = (-0.2, 0.3), the three boundary components are $B_{5,g} = 0.838$, $B_{8,g} = 0.425$, and $B_{8,a} = 0.239$. Their arithmetic mean is 0.501, whereas the harmonic mean is 0.389. The harmonic mean therefore implements the intended conservative requirement that cross-scale boundary evidence should persist rather than be supplied by only one view.

### 8.8 Baseline comparison with nearest-document distance maximization

To test whether the Supported Knowledge Gap formulation provides information beyond a simple empty-space criterion, we evaluated a nearest-document-distance baseline on the same 21 × 21 grid. The baseline selected the 16 locations with the largest nearest-document distance $d_{nn}$, matching the size of the practical KMO front. These baseline points were substantially farther from observed documents than the KMO solutions (mean $d_{nn} = 0.211$ versus 0.117), but all 16 lay on the outer boundary of the fixed chart. Their mean evidential support was only S = 0.086 compared with 0.828 for KMO, while mean reliability fell to R = 0.001 compared with 0.761. Persistent-boundary strength and semantic gradient also decreased markedly ($P_{boundary} = 0.382$ versus 0.856; G = 0.008 versus 0.842). Consistent with this loss of support, the mean GPR uncertainties increased from 0.222 to 0.374 for the RBF model and from 0.340 to 0.504 for the Matérn-3/2 model.

Because the unconstrained distance baseline preferentially selected chart-edge locations, the comparison was repeated after excluding all grid points on x = ±1 or y = ±1. This interior-only distance baseline still produced substantially weaker candidates: mean $d_{nn} = 0.151$, S = 0.309, R = 0.401, $P_{boundary} = 0.444$, and G = 0.354. Its mean RBF and Matérn uncertainties were 0.278 and 0.393, respectively. Thus, the difference is not solely a consequence of the bounded-chart geometry. Maximizing geometric emptiness alone tends to move the search away from evidential support and meaningful semantic structure, whereas KMO explicitly seeks underrepresented regions that remain supported by neighboring literature.

Table 4. Baseline comparison of mean discovery and support metrics for 16 selected locations.

| Method | mean $d_{nn}$ | mean S | mean R | mean $P_{boundary}$ | mean G |
|---|---|---|---|---|---|
| KMO practical front | 0.117 | 0.828 | 0.761 | 0.856 | 0.842 |
| Distance-max, all grid | 0.211 | 0.086 | 0.001 | 0.382 | 0.008 |
| Distance-max, interior only | 0.151 | 0.309 | 0.401 | 0.444 | 0.354 |

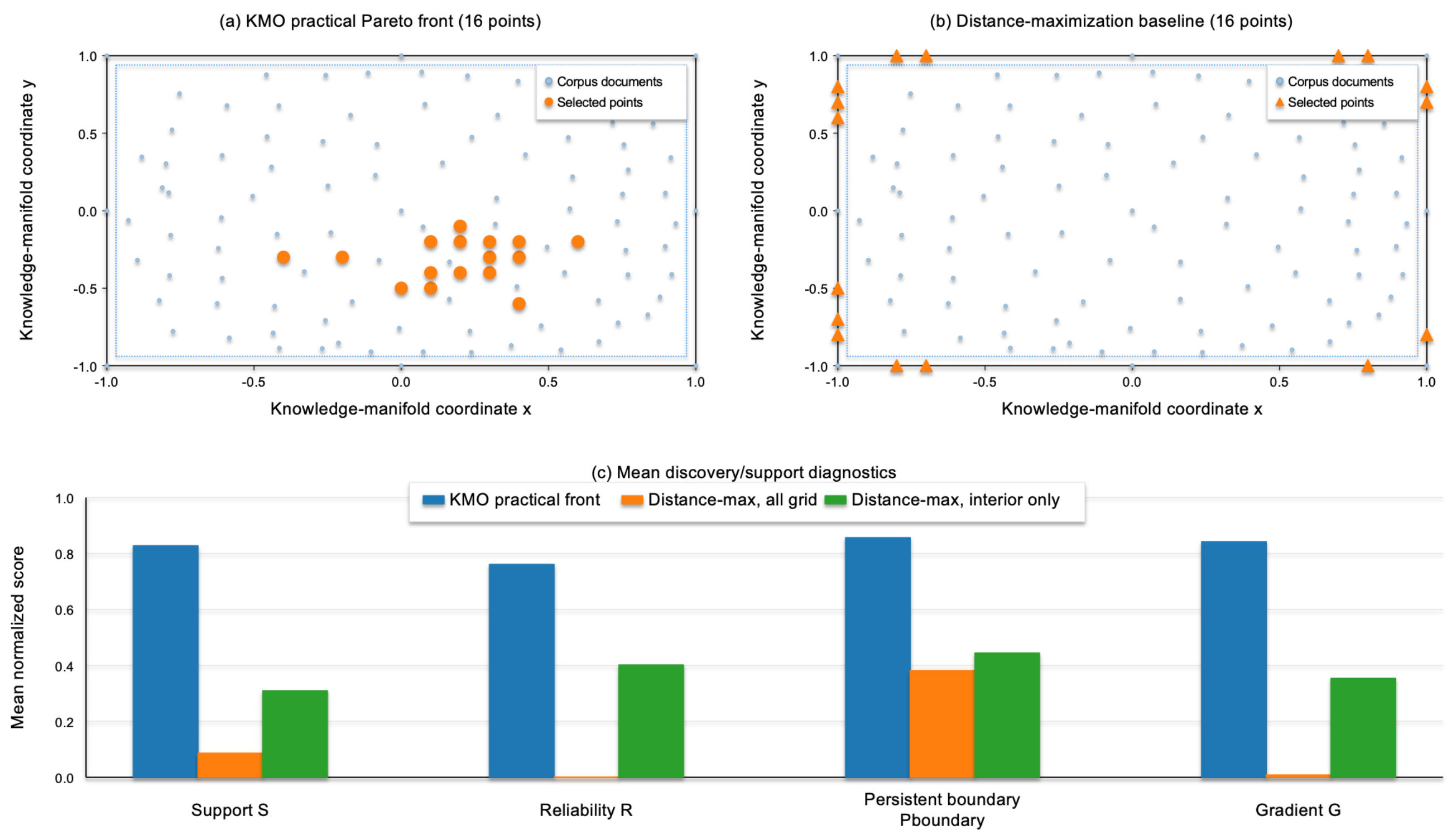

Figure 8. Baseline comparison between the 16 practical KMO Pareto solutions and a simple nearest-document-distance maximization baseline. Panels (a) and (b) show the selected locations on the same fixed Knowledge Manifold. All 16 unconstrained distance-maximization points lie on the chart boundary. Panel (c) compares mean support S, reliability R, persistent-boundary score $P_{boundary}$, and semantic-gradient score G for KMO, the unconstrained distance baseline, and an interior-only distance baseline. The results show that maximum emptiness preferentially selects weakly supported, high-uncertainty extrapolative regions rather than Supported Knowledge Gaps. Note that the mean reliability of the all-grid distance-maximization baseline is R = 0.001; the corresponding bar in panel (c) is therefore nearly invisible at the plotted scale.

### 8.9 Scope of robustness claims

The sensitivity analyses support three conclusions. First, the main discovery regions are not consequences of a single uncertainty kernel or a single clustering resolution. Second, local interpretation is less stable than broad map structure, making continuous fields preferable to hard cluster labels. Third, the Pareto framework is more informative than a single scalar score because it retains multiple scientifically interpretable discovery mechanisms.

These tests do not, however, establish external scientific novelty. The present analysis demonstrates robustness within the constructed 99-paper aerodynamic-optimization Knowledge Manifold. Whether a candidate Supported Knowledge Gap corresponds to genuinely unpublished knowledge must be evaluated separately using broader literature retrieval and domain-expert judgment.

## 9. Discussion

### 9.1 From design-space exploration to knowledge-space exploration

The principal conceptual contribution of this study is the extension of multi-objective design exploration from an engineering design space to a scientific knowledge space.

In our earlier aerodynamic and multidisciplinary design studies, Pareto-optimal solutions were not regarded merely as alternative optimum designs. By projecting and analyzing the Pareto set using self-organizing maps and related data-mining techniques, relationships among objectives, design variables, and characteristic design regimes could be extracted as design knowledge [30–33]. This viewpoint ultimately motivated Multi-Objective Design

Exploration (MODE), in which optimization serves not only to identify high-performance solutions but also to reveal the structure of the underlying design space.

Knowledge Manifold Optimization transfers the same principle to scientific literature. The analogy is not exact, but it is operationally useful. The important commonality is that the set of non-dominated solutions carries information that is lost when one solution is selected prematurely. In engineering design, different Pareto solutions represent different trade-offs among performance criteria. In the present problem, different Pareto locations represent different mechanisms of potential scientific discovery.

This explains why the practical Pareto front contains several distinct discovery archetypes rather than converging to one preferred point. The adjoint/methodological bridge, aero-structural topology/FSI region, functional aerodynamic-device region, and rear-wake/vehicle flow-control region correspond to different compromises among boundary persistence, novelty, semantic change, and evidential support. The diversity of these solutions is therefore not an inconvenience of multi-objective optimization; it is part of the information being sought.

Table 5. Conceptual correspondence between multi-objective engineering design exploration and Knowledge Manifold Optimization.

| Engineering design | Scientific knowledge |
| --- | --- |
| design variables | manifold coordinates |
| performance objectives | discovery objectives |
| Pareto-optimal designs | Pareto-optimal knowledge gaps |
| SOM / design-space analysis | Knowledge Manifold analysis |
| design knowledge | candidate scientific knowledge |

## 9.2 What constitutes a Supported Knowledge Gap?

A central result of the present study is that empty space alone is an insufficient definition of a knowledge gap.

If novelty were equated with distance from existing documents, optimization would tend toward sparsely populated regions at the edge of corpus support. Such locations may appear novel geometrically but can lack sufficient evidence for meaningful interpretation. Conversely, a highly supported point in a dense literature region may offer little beyond incremental variation. The explicit distance-maximization baseline in Fig. 8 confirms this behavior quantitatively: maximizing nearest-document distance alone selects low-support, high-uncertainty locations, even when chart-boundary points are excluded.

The band-pass white-space objective formalizes this distinction by assigning the highest discovery value to intermediate distances from existing documents. Novelty, however, is only one dimension. The strongest candidate regions also tend to occur where several knowledge regimes compete and where the semantic field varies substantially. This suggests that useful scientific gaps are not simply vacancies but structured interfaces between existing bodies of knowledge.

The term Supported Knowledge Gap was introduced to capture this distinction. Such a point is not required to be maximally distant from existing work. Rather, it occupies a region that is sufficiently underrepresented to offer novelty, sufficiently close to multiple existing works to retain evidential support, and sufficiently heterogeneous to permit a meaningful synthesis of neighboring knowledge.

This interpretation is consistent with the broader idea of literature-based discovery, in which potentially valuable knowledge can arise from connections that are individually supported but have not previously been combined explicitly [24,25]. Knowledge Manifold Optimization adds a continuous geometric formulation to this idea: rather than searching only for discrete term chains or citation-network bridges, it evaluates every candidate location within a fixed knowledge space.

too close → incremental | intermediate → supported novelty | too far → unsupported extrapolation

### 9.3 Why scientific discovery is treated as a multi-objective problem

The results also show why a single scalar discovery score is difficult to justify. The exploratory six-objective formulation produced 139 non-dominated points among 441 candidates, demonstrating that excessive dimensionality makes Pareto dominance weakly selective. Conversely, assigning fixed weights to all discovery indicators produces an apparently simple ranking but embeds subjective assumptions about how different types of scientific value should compensate for one another.

For example, there is no natural scientific conversion factor stating how much additional semantic gradient should compensate for a reduction in evidential support, or how much additional white space should compensate for weaker cross-scale boundary persistence.

The final formulation therefore retains four quantities with distinct interpretations: $P_{boundary}$, W, G, and S. Reliability and cross-model uncertainty consistency are treated separately as diagnostics. This separation is conceptually important. Reliability answers how cautiously a candidate should be interpreted, whereas the four Pareto objectives answer why it may be scientifically interesting.

The resulting practical front of 16 locations is sufficiently small to permit interpretation while remaining diverse enough to preserve different discovery mechanisms. The point (0.3,−0.3), for example, is a balanced compromise under an equal-distance-to-ideal criterion, but the framework deliberately does not call it the best discovery point. Choosing it over other Pareto solutions would require an additional preference model that is external to the present analysis.

### 9.4 Multi-resolution boundaries as evidence of interdisciplinarity

One of the most useful aspects of the present analysis is the distinction between a boundary that is visible only at one resolution and a boundary that persists across different representations.

A location may appear interdisciplinary under broad global smoothing simply because multiple distant research regimes are averaged together. The adaptive K = 8 representation provides a stricter local test. Several initially attractive points collapsed toward one dominant research regime when examined locally, whereas the strongest candidates at (0.2,−0.1) and (0.1,−0.5) retained substantial contributions from multiple regimes.

The harmonic-mean boundary objective was introduced specifically to penalize such scale-dependent artifacts. Because the harmonic mean is non-compensatory relative to the arithmetic mean, a weak boundary signal in any one of the three views substantially lowers the combined score. This formulation therefore treats cross-scale boundary character as a persistence property rather than as the result of a single clustering or smoothing choice.

The choice of K = 5 and K = 8 should nevertheless not be interpreted as evidence that the aerodynamic literature possesses exactly five or eight objectively defined fields. The clusterings are analytical resolutions imposed on a continuous semantic landscape. In the present corpus, K = 5 provided a useful wide-angle partition and K = 8 provided a useful detailed reference. Their scientific value lies in the complementary structure they reveal, not in a claim of uniquely correct taxonomy.

### 9.5 Relation to recent AI-based scientific discovery

Knowledge Manifold Optimization occupies a different position from recent LLM-centered scientific-discovery systems [35–39]. Systems such as SciMON and ResearchAgent use scientific literature to help generate or refine research ideas. AUTODISCOVERY goes further by explicitly treating discovery as search and optimizing Bayesian surprise within an LLM-generated hypothesis space. Mechanism-centric cross-domain retrieval similarly attempts to identify potentially fruitful connections beyond straightforward semantic similarity.

These approaches and the present framework share an important premise: scientific discovery can be made, at least in part, into a computational search problem. The principal difference lies in what is searched.

In LLM-centered approaches, the search space is typically composed of generated hypotheses, candidate experiments, retrieved inspirations, or reasoning branches. In Knowledge Manifold Optimization, the search space is constructed before hypothesis generation as an explicit geometrical representation of the literature: literature → knowledge space → numerical search → hypothesis.

This ordering is intentional. It allows the location of a candidate, its supporting papers, its local regime mixture, its semantic gradient, and its uncertainty to be inspected independently of the language model that eventually describes it.

The approaches should therefore be considered complementary rather than competing. Bayesian surprise, mechanistic similarity, or experimental utility could in the future be added as downstream validation criteria or even additional discovery objectives, while Knowledge Manifold Optimization provides an explicit geometrical substrate on which candidate regions are first identified.

## 9.6 The role of the language model

The language model in the present framework is deliberately assigned a narrower role than in autonomous scientific-discovery systems.

The LLM does not construct the Knowledge Manifold, calculate the discovery fields, choose the objective functions, identify the Pareto front, or select the coordinates of candidate gaps. These operations are performed numerically.

Only after a candidate has been identified is an evidence package generated containing local semantic terms, document contributions, document frequencies, support measures, knowledge-regime probabilities, and uncertainty. The LLM then converts that evidence into a human-readable candidate research description.

These constraints do not constitute a formal guarantee against hallucination or overinterpretation. Instead, they make unsupported statements identifiable and auditable because the generated text can be compared directly with the finite evidence package provided to the model. Rare terms are particularly important in this respect: terms appearing in only one or a few neighboring documents are not promoted automatically to regional knowledge, but must either be identified explicitly as limited-source hypotheses or be excluded from the synthesis. Appendix A provides the complete researcher-supplied prompt and output-audit protocol, and Appendix B provides the two full virtual abstracts used as representative examples.

Thus, the generated virtual paper is better interpreted as an evidence-constrained hypothesis proposal than as an autonomous scientific discovery.

geometry determines where + evidence determines what + LLM determines how to express it

## 9.7 Corpus-relative novelty and external scientific novelty

The strongest limitation of the present study is also conceptually important: the method identifies novelty relative to the analyzed corpus.

The 99-paper dataset was created by first obtaining 150 papers recommended by LeapSpace for aerodynamic optimization over the preceding ten years, then retaining papers for which PDF files could be downloaded through Scopus and excluding review articles. This provides a focused and practically usable corpus, but it is not an exhaustive sample of all aerodynamic-optimization literature.

Three forms of selection therefore affect the final manifold: the initial recommendation mechanism, the ten-year temporal window, and full-text availability. Excluding review papers also changes the semantic structure intentionally by removing documents whose purpose is to synthesize large parts of the field.

Consequently, Supported Knowledge Gap is not equivalent to a globally novel scientific topic. The former is a computational result; the latter is an external scientific judgment.

A promising next stage is therefore a two-step validation protocol: Knowledge Manifold candidate → external literature search → domain-expert evaluation. A candidate that survives both stages would provide considerably stronger evidence of genuine research opportunity.

The corpus size should also be distinguished from the size of the optimization problem. The 99-document corpus limits the empirical scope of the present scientific conclusions, but it does not impose a corresponding limit on the number of candidate locations that can be searched by KMO. Conversely, scaling the literature corpus itself requires approximate manifold-construction and uncertainty-estimation methods. Thus, corpus coverage and computational scalability are separate issues.

## 9.8 Methodological limitations

Several additional limitations should be noted.

First, the two-dimensional Knowledge Manifold is an operational reduced representation and should not be interpreted as proof that scientific knowledge has intrinsic dimension two. The measured stress, distance correlations, trustworthiness, and continuity indicate useful global and mesoscale preservation, but local neighbor relations are not exact.

Second, the map depends on the document representation and gauge-fixing procedure. The character n-gram TF–IDF representation was selected partly because of its interpretability and auditability, but alternative dense representations could produce different semantic geometries. Likewise, changing the anchor configuration affects the map sufficiently that the fixed chart should be regarded as part of the model specification.

Third, cluster structure is not unique. The variation observed across $K = 3$–$9$ and across initialization demonstrates that hard cluster labels should not be regarded as intrinsic scientific categories. This is why the final formulation emphasizes continuous mixture fields and boundary persistence.

Fourth, the discovery objectives themselves are designed quantities. Although robust percentile normalization and sensitivity analyses reduce dependence on arbitrary scales, the precise form of the entropy/gap combination, the white-space band, and the support function remains heuristic. Their purpose is to express transparent scientific preferences, not to claim that they constitute a uniquely correct mathematical definition of discovery.

Fifth, the present Pareto search is conducted on a $21 \times 21$ grid. Exact dominance testing therefore identifies the complete Pareto set only with respect to these 441 sampled locations. Finer grids or continuous multi-objective optimization could reveal additional solutions.

Finally, the four discovery archetypes are a post hoc interpretive typology rather than algorithmically defined clusters. As an independent geometric reference, Ward hierarchical clustering in both objective space and manifold-coordinate space produced partitions that partially agreed with the semantic grouping. This partial correspondence provides geometric support for the typology while also showing that local semantic evidence, contributing-document provenance, and multi-resolution mixture structure contain information not captured by Euclidean proximity alone. The archetypes should therefore be regarded as an explanatory summary of the present Pareto set, not as a unique or intrinsic partition of the discovery landscape.

## 9.9 Future directions

Several extensions follow naturally from the present formulation. The most direct is continuous Knowledge Manifold Optimization, in which the manifold coordinates themselves become continuous decision variables. Multi-objective optimizers such as NSGA-II could then search $x \in [-1,1]^2$ without dependence on a fixed evaluation grid.

A second direction is expansion from a single 99-paper aerodynamic corpus to substantially larger and more diverse corpora. Cross-domain tests would be particularly valuable because the method explicitly targets boundaries between knowledge regimes.

A third direction is external validation. Candidate Supported Knowledge Gaps could be subjected to systematic literature searches and blinded expert assessment of novelty, plausibility, and potential scientific value. Such evaluation would allow the discovery objectives themselves to be calibrated empirically.

A fourth possibility is to incorporate additional information into the objective system. Citation-network structure, temporal emergence, experimental feasibility, mechanistic similarity, Bayesian surprise, and expected scientific utility could complement the present geometric objectives.

Finally, the framework could be made longitudinal. If a Knowledge Manifold is frozen at time $t$, candidate gaps identified at that time can later be compared with publications appearing at $t + \Delta t$. Such a prospective experiment would provide a particularly strong test: whether regions identified as promising gaps are disproportionately occupied by subsequent scientific work.

In that sense, the ultimate validation of Knowledge Manifold Optimization would not be whether a generated virtual paper sounds plausible, but whether the geometry identifies future scientific activity—or genuinely productive research directions—better than appropriate baseline strategies.

## 10. Conclusions

This study introduced Knowledge Manifold Optimization, a discovery-oriented extension of the Knowledge Manifold framework. Previous work established a continuous, differentiable, and uncertainty-aware representation of scientific literature in which semantic content, gradients, uncertainty, and evidence can be evaluated at locations not occupied by existing documents [28,29]. The present study addressed the subsequent question: where within such a knowledge space should one search for potentially new scientific knowledge?

Scientific discovery was formulated as a multi-objective problem rather than as maximization of empty space or of a single weighted discovery score. Four complementary objectives were used: $P_{boundary}$, W, G, and S, representing persistent knowledge-regime boundaries, band-pass white-space novelty, semantic-gradient magnitude, and evidential support, respectively. Model uncertainty was retained as a separate reliability measure rather than incorporated directly into the discovery objective.

The framework was applied to a corpus of 99 aerodynamic-optimization papers assembled from an initial set of 150 LeapSpace recommendations from the preceding ten years, subject to full-text availability through Scopus and exclusion of review articles. Multi-resolution analysis used a broad K = 5 discovery representation and a detailed K = 8 reference representation, together with global and adaptive interpolation to test whether apparent knowledge boundaries persisted across resolution and locality.

Among 441 evaluated manifold locations, the final four-objective formulation produced 19 non-dominated points, of which 16 constituted the practical Pareto front after requiring nonzero white-space support. These points did not represent one unique optimum. Instead, they revealed several distinct discovery archetypes involving methodological bridges, aero-structural topology and fluid–structure interaction, functional aerodynamic devices and flow phenomena, and rear-wake/vehicle flow-control integration. A matched-size distance-maximization baseline selected substantially more remote but much less supported and less reliable locations, demonstrating that maximum emptiness is not equivalent to a Supported Knowledge Gap.

Two representative Pareto solutions further illustrated different mechanisms of discovery. The location (0.2,−0.1) combined strong semantic variation with functional aerodynamic concepts such as attachment, circulation, and vortex behavior, whereas (0.1,−0.5) provided a highly persistent bridge among topology-oriented material design, structural mechanics, aerodynamics, and fluid–structure interaction. Evidence packages extracted at these locations were subsequently verbalized into candidate research hypotheses, with the language model acting only after the numerical search. The complete prompt-level constraints and the full generated examples are documented in Appendices A and B.

The conceptual origin of this approach is closely related to Multi-Objective Design Exploration. In earlier engineering applications, Pareto-optimal design sets were analyzed using self-organizing maps to extract design knowledge. Knowledge Manifold Optimization extends the same principle from Pareto solutions → design knowledge to Pareto knowledge gaps → candidate scientific knowledge.

The term Supported Knowledge Gap is intentionally corpus-relative. The present method identifies regions that are underrepresented yet supported within the analyzed literature; it does not by itself establish global scientific novelty. Broader literature search, expert evaluation, and eventually prospective validation are required before such candidates can be regarded as genuine discoveries.

The principal implication of this study is therefore methodological rather than domain-specific: scientific discovery can be approached not only as the generation of hypotheses, but also as a structured multi-objective search over an explicit knowledge space. Knowledge Manifold Optimization provides one framework for carrying out that search while keeping the origin, support, uncertainty, and interpretation of each candidate auditable.

## Appendix A. Evidence-Constrained LLM Verbalization Protocol

This appendix documents the complete researcher-supplied task prompt used for evidence-constrained verbalization. It is the reproducible research prompt supplied together with the machine-generated evidence package; it should not be confused with any provider-level or proprietary system instruction. The prompt is reproduced verbatim below.

### A.1 Prompt template (verbatim)

```
You are participating in a controlled experiment.  A deterministic
pipeline has analyzed a corpus of papers and exported a machine-computed
**evidence package** for one location on its knowledge map — the attached
`evidence_point_<x>_<y>.json`, as produced by `code/make_evidence.py`.
No real paper occupies that location.  Your task is to write the
**virtual document**: a description of the research that *would* occupy
it.

Evidence terms are **whole words**, deterministically reconstructed from
the corpus text (each entry lists its attested `word_forms`).  No
fragment decoding is needed.

The package contains:
- `theme_lens_L2` — words indicating the local research theme.
- `concentration_lens_L1` — words whose probability mass is locally
  concentrated; each carries `df`, the number of corpus documents
  containing the word, and, where `df <= 3`, its `source_docs`.
- `contributing_documents_top3` — the highest-weighted neighboring
  papers (titles and weights).
- `mixture_entropy_H`, `effective_contributing_documents`,
  `relative_uncertainty_u` — how broad and how well-supported the
  location is.

Rules (binding)

1. **Theme from the theme lens.**  Build the consensus research theme
   from the L2 words and the contributing documents.  Words appearing
   in BOTH lens lists carry the greatest weight.  Titles of
   contributing documents may help you interpret the words, but the
   description must not be a paraphrase of those titles.
2. **Serendipity from the concentration lens, graded by df.**  Words
   appearing only in the L1 list are candidates for speculative
   additions.  If `df >= 4`: you may present the word as an emergent
   shared topic.  If `df <= 3`: you may mention it ONLY as a
   single-source hypothesis, explicitly citing its `source_docs`
   (e.g. "as explored in only one neighboring study [doc 18]").
   Never present a `df <= 3` word as established regional knowledge.
3. **No outside facts.**  Do not assert anything the evidence package
   does not support.  Your general knowledge of composites may be used
   to connect the evidence into fluent prose, not to add claims absent
   from it.
4. **Use every listed word or account for it.**  All terms in the
   package are corpus words that survived a noise filter; none is a
   bibliographic fragment.  If you nevertheless judge a term
   unusable, say so explicitly in one line rather than silently
   dropping it.
5. **Close with the evidence list.**  End your answer with a section
   `Evidence used:` listing every word you relied on, each with its
   df, plus any term you set aside under Rule 4.

Deliverables

- **Task A**: the virtual document, in English, 200–280 words,
  structured as a research abstract (topic, approach, expected
  contribution).  Include a one-line caveat reflecting
  `relative_uncertainty_u` and `effective_contributing_documents`.
- **Task B**: a 45-word summary of the same virtual document.

Output Task A, then Task B, then the `Evidence used:` list.  Do not
include anything else — no reasoning trace, no extra sections.
```

### A.2 Post-generation audit protocol

The two representative outputs reported in Appendix B were checked against the following audit items. The audit is intended to expose unsupported extrapolation rather than to claim a mathematical guarantee of hallucination-free generation.

- Length compliance: Task A contains 200–280 words and Task B contains exactly 45 words.
- Evidence coverage: every supplied semantic term is either used in the synthesis or explicitly identified as set aside.
- Lens discipline: L2 thematic terms define the principal topic; L1 terms are treated as supporting or concentration-sensitive evidence.
- Low-frequency provenance: every L1-only term with df $\leq 3$ is presented only as a limited-source hypothesis and its source document is identified explicitly.
- Unsupported-claim audit: technical claims, mechanisms, materials, numerical performance claims, and citations not supported by the evidence package are not introduced as facts.
- Novelty wording: the output does not claim global scientific novelty; novelty remains relative to the analyzed corpus unless externally validated.
- Uncertainty disclosure: relative uncertainty and effective contributing-document support are stated explicitly in the abstract.
- Evidence trace: the output ends with an Evidence used list that records the terms used and their document frequencies, including any term set aside.

For the examples below, Task A contains 245 words in each case and Task B contains exactly 45 words in each case. The audit therefore verifies format compliance while preserving the more important distinction between evidence-supported statements and hypotheses requiring external validation.

## Appendix B. Full Examples of Evidence-Constrained Virtual Research Abstracts

The following outputs are reproduced in full to make the endpoint of Knowledge Manifold Optimization explicit. They are virtual research documents generated from corpus-derived evidence packages, not claims that the proposed studies are externally novel, feasible, or scientifically validated.

### B.1 Candidate C: functional aerodynamic integration at $x = (0.2, -0.1)$

Representative Pareto candidate from Archetype C. The evidence package emphasizes rear-element aerodynamics, airflow attachment, vortex/circulation behavior, operating angles, and a carefully limited treatment of lower-frequency terms.

**Task A — Virtual research abstract (245 words)**

This virtual study investigates how aerodynamic attachment, airflow, vortex organization, and circulation interact around rear lifting or control features across varying angles, with particular attention to configurations in which local flow organization enhances overall aerodynamic behavior. The proposed approach uses a parameterized optimization study to vary rear-element geometry and operating angles, then examines how changes in attachment and reattachment, vortical structure, and recirculating flow are associated with performance trends. Because both the theme and concentration lenses emphasize rear, airflow, attachment, angles, vortex, enhances, and circulation, the central hypothesis is that coordinated control of these features can identify design regions where favorable circulation and persistent attachment emerge together, indicating a robust aerodynamic mechanism rather than an isolated response.
The concentration lens additionally suggests biplane geometry as an emergent shared topic (df=4), so a biplane-type arrangement is included as a secondary design family rather than treated as the dominant regional theme. A Trefftz-related interpretation is considered only as a single-source hypothesis, because this term appears in one neighboring study [doc 89]; it is therefore used only as a possible analysis perspective, not as established local knowledge. The expected contribution is a unified design framework linking rear-element placement, operating angles, airflow attachment, vortex behavior, and circulation to identify configurations that enhance performance while remaining interpretable across several neighboring aerodynamic studies.
Caveat: relative uncertainty is 0.222 with 73.6 effective contributing documents, indicating broad support but not direct evidence for a real paper at this map location.

**Task B — 45-word summary**

This virtual study explores optimization of rear aerodynamic elements across varying angles, linking airflow attachment, vortex behavior, and circulation to enhanced performance. It also considers biplane configurations and a cautiously framed Trefftz perspective, while emphasizing broadly supported patterns over single-source evidence at the knowledge-map location.

**Evidence used**

- rear — df=25 (L2); df=25 (L1) — used.
- airflow — df=24 (L2); df=24 (L1) — used.
- attachment — df=24 (L2); df=21 (L1) — used, including the reattachment/detachment family.
- enhances — df=32 (L2); df=32 (L1) — used.
- angles — df=44 (L2); df=44 (L1) — used.
- vortex — df=57 (L2); df=57 (L1) — used, including vortices/vorticity/vortical forms.
- indicating — df=51 (L2) — used as an evidence-linked descriptive term.
- circulation — df=40 (L2); df=40 (L1) — used, including recirculation/recirculating/circulating forms.
- biplane — df=4 (L1) — used as an emergent shared topic, not as the dominant theme.
- trefftz — df=1 (L1) — used only as a single-source hypothesis from [doc 89].
- program — df=54 (L1) — set aside because the reconstructed forms are semantically heterogeneous and do not support a specific technical claim here.

## B.2 Candidate B: aero-structural topology and FSI at x = (0.1, −0.5)

Representative Pareto candidate from Archetype B. The evidence package emphasizes topology-oriented material design, compliance and stiffness, structural behavior, penalization, and fluid–structure interaction.

**Task A — Virtual research abstract (245 words)**

This virtual study investigates a topology-oriented material design framework for aerodynamic systems in which structural response and fluid-structure effects are considered together. The local theme is defined most strongly by material, topology, compliance, stiffness, structural, and penalization, while the nearest contributing studies indicate a surrounding context of aerodynamic optimization spanning conceptual configuration design, regime-dependent optimization, and robust high-fidelity shape optimization.

The proposed approach would represent candidate designs through topology changes in one- or multi-material layouts, while using compliance and stiffness as complementary indicators of structural behavior. Penalization would be incorporated as a central mechanism for distinguishing preferred material distributions during optimization. The study would then examine how structural and fluid-structure considerations alter the resulting topologies, with particular attention to whether optimized layouts develop holes or connected material regions as recurring geometric outcomes. The concentration lens also retains the term Sigmund, which is treated here as an emergent methodological cue within the local corpus rather than as evidence for any single specific formulation.

The expected contribution is a unified virtual research direction linking aerodynamic design, structural topology, material allocation, compliance-stiffness tradeoffs, penalization, and fluid-structure interaction within one optimization setting. Such a study would occupy the gap between primarily aerodynamic optimization and structurally informed topology design, while remaining anchored to the surrounding evidence rather than extrapolating to an unsupported application.

Caveat: the location has relative uncertainty u = 0.23 and an effective contribution from 72.0 documents, indicating broad support but not a uniquely determined research interpretation.

**Task B — 45-word summary**

This virtual study proposes topology-oriented material optimization for aerodynamic systems, combining compliance, stiffness, structural response, penalization, and fluid-structure effects. It explores material layouts and holes within optimized topologies, linking aerodynamic design with structurally informed optimization while remaining supported by a broad neighboring-document mixture evidence base.

**Evidence used**

- material — df = 49
- topology — df = 45
- compliance — df = 17
- stiffness — df = 20
- structural — df = 61
- penalization — df = 24
- fluid-structure — df = 18
- sigmund — df = 18
- holes — df = 15